\documentclass[letterpaper,11pt]{article}
\usepackage{url}
\usepackage{amssymb,amsmath,amsfonts,amsthm}
\usepackage{fullpage}
\usepackage[usenames]{color}

\newcommand{\polylog}{{\mathrm{polylog}}}
\newcommand{\hash}{\mathcal{H}}
\newcommand{\sign}{\mathrm{sign}}

\DeclareMathSymbol{\qedsymb} {\mathord}{AMSa}{"04}

\newcommand{\eps}{\varepsilon}

\newcommand{\indicate}{\mathbf{1}}

\newcommand{\prefix}{\mathsf{prefix}}
\newcommand{\fphh}{\mathsf{F_pHH}}
\newcommand{\basicfphh}{\mathsf{BasicF_pHH}}
\newcommand{\fpest}{\mathsf{F_pEst}}
\newcommand{\query}{\mathsf{Query}}
\newcommand{\countsketch}{\mathsf{CountSketch}}
\newcommand{\countmin}{\mathsf{CountMin}}
\newcommand{\highend}{\mathsf{HighEnd}}
\newcommand{\filter}{\mathsf{Filter}}
\newcommand{\basichighend}{\mathsf{BasicHighEnd}}

\newcommand{\lightest}{\mathsf{LightEstimator}}
\newcommand{\Est}{\mathsf{Est}}

\newcommand{\fpreport}{\mathsf{F_pReport}}
\newcommand{\fpupdate}{\mathsf{F_pUpdate}}
\newcommand{\fpspace}{\mathsf{F_pSpace}}

\newcommand{\inprod}[1]{\langle #1 \rangle}

\newcommand{\ceil}[1]{\left\lceil #1 \right\rceil}

\newcommand{\oct}{\quad\quad}                                   

\newcommand{\x}{x}

\newcommand{\R}{\mathbb{R}}
\newcommand{\C}{\mathbb{C}}

\newcommand{\Real}{\mathbf{Re}}
\newcommand{\Imag}{\mathbf{Im}}

\newcommand{\E}{\mathbf{E}}

\newcommand{\Var}{\mathbf{Var}}
\renewcommand{\Pr}{\mathbf{Pr}}

\newcommand{\tO}{\tilde{O}}

\newcommand{\BoldTheorem}[1]{\textbf{Theorem~\ref{thm:#1} (restatement).}}

\newcommand{\DefinitionName}[1]{\label{def:#1}}
\newcommand{\EquationName}[1]{\label{eq:#1}}

\newcommand{\LemmaName}[1]{\label{lem:#1}}

\newcommand{\RemarkName}[1]{\label{rem:#1}}
\newcommand{\SectionName}[1]{\label{sec:#1}}
\newcommand{\TheoremName}[1]{\label{thm:#1}}
\newcommand{\FigureName}[1]{\label{fig:#1}}

\newcommand{\Equation}[1]{Eq.\:\eqref{eq:#1}}

\newcommand{\Lemma}[1]{Lemma~\ref{lem:#1}}

\newcommand{\Remark}[1]{Remark~\ref{rem:#1}}
\newcommand{\Section}[1]{Section~\ref{sec:#1}}
\newcommand{\Theorem}[1]{Theorem~\ref{thm:#1}}
\newcommand{\Figure}[1]{Figure~\ref{fig:#1}}

\newtheorem{theorem}{Theorem}

\newtheorem{definition}[theorem]{Definition}

\newtheorem{lemma}[theorem]{Lemma}
\newtheorem{remark}[theorem]{Remark}

\newcommand{\proofbelow}{3pt}
\newcommand{\afterproof}{\hfill $\blacksquare$ \par \vspace{\proofbelow}}
\newcommand{\aftersubproof}{\hfill $\Box$ \par \vspace{\proofbelow}}
\renewenvironment{proof}{\noindent\textbf{Proof.}\,}{\afterproof}

\newcommand{\poly}{\mathop{{\rm poly}}}
\renewcommand{\th}{\ifmmode{^{\textrm{th}}}\else{\textsuperscript{th}\ }\fi}

\newcommand{\eqdef}{\mathbin{\stackrel{\rm def}{=}}}
\newcommand{\comment}[1]{}

\begin{document}

\author{Daniel M. Kane\footnotemark[2]\oct
Jelani Nelson\footnotemark[3]\oct
  Ely Porat\footnotemark[4]\oct
  David P. Woodruff\footnotemark[5]
  }

\date{}

\footnotetext[1]{Harvard University, Department of
  Mathematics. \texttt{dankane@math.harvard.edu}.}
\footnotetext[2]{MIT Computer Science and Artificial Intelligence
  Laboratory. \texttt{minilek@mit.edu}. 
}
\footnotetext[3]{Department of Computer Science, Bar Ilan
  University. \texttt{porately@macs.biu.ac.il}.}
\footnotetext[4]{IBM Almaden Research Center, 650 Harry Road, San
  Jose, CA, USA. \texttt{dpwoodru@us.ibm.com}.}

\title{Fast Moment Estimation in Data Streams in Optimal Space}

\maketitle

\begin{abstract}
We give a space-optimal algorithm with update time
$O(\log^2(1/\eps)\log\log(1/\eps))$
for $(1\pm\eps)$-approximating the $p$th
frequency moment, $0 < p < 2$, of a length-$n$ vector updated in a
data stream.  This provides a
nearly exponential improvement in the update time complexity over the
previous space-optimal algorithm of [Kane-Nelson-Woodruff, SODA 2010],
which had update time
$\Omega(1/\eps^2)$. 
\end{abstract}


\section{Introduction}\SectionName{intro}
The problem of estimating frequency moments of a vector being updated in
a data stream was first studied by
Alon, Matias, and Szegedy \cite{AMS99} and has since received much
attention
\cite{BJKS04,bgks06,Indyk06,IW05,KNW10b,Li08b,NW10,W04,WoodruffThesis}.
Estimation of the second moment has applications to estimating join
and self-join sizes \cite{AlonGMS02} and to network anomaly detection
\cite{KSZC03,ThorupZhang04}. First moment estimation is useful in
mining network traffic data \cite{CMR05}, comparing empirical
probability distributions, and several other applications (see
\cite{NW10} and the references therein). Estimating fractional moments
between the $0$th and $2$nd has applications to entropy estimation for
the purpose of network anomaly detection \cite{HNO08, ZhaoLOSWX07},
mining tabular data \cite{CormodeIKM02}, image decomposition
\cite{GeigerLD99}, and weighted sampling in
turnstile streams \cite{MW10}. It was also observed
experimentally that the use of fractional moments in
$(0,1)$ can improve the effectiveness of standard clustering
algorithms \cite{ahk01}.

Formally in this problem, we are given up front a real number $p>0$.
There is also an underlying $n$-dimensional
vector $x$ which starts as $\vec{0}$. What follows is a sequence of
$m$ updates of the form $(i_1,v_1),\ldots,(i_m,v_m)\in [n]\times
\{-M,\ldots,M\}$ for some $M>0$. An update $(i,v)$ causes the change
$x_i\leftarrow x_i + v$.  We would like to compute $F_p \eqdef
\|x\|_p^p \eqdef \sum_{i=1}^n |x_i|^p$, also called the {\em $p$th
  frequency moment} of $x$.  In many applications, it is
required that the algorithm only use very limited space while processing the
stream, e.g., in networking applications where $x$ may be indexed by
source-destination IP pairs and thus a router cannot afford to store
the entire vector in memory, or in database applications where one
wants a succinct ``sketch'' of some dataset, which can be
compared with short sketches of other datasets for fast computation of
various (dis)similarity measures.

Unfortunately, it is known
that linear space ($\Omega(\min\{n,m\})$ bits) is required unless one
allows for
(a) {\em approximation}, so that we are only guaranteed to output a
value in $[(1-\eps)F_p, (1+\eps)F_p]$ for some $0<\eps<1/2$, and (b)
{\em randomization}, so that the output is only guaranteed to be
correct
with some probability bounded away from $1$, over the randomness used
by the algorithm \cite{AMS99}.  Furthermore, it is known that
polynomial space is
required for $p>2$ \cite{BJKS04,cks03,Gronemeier09,Jayram09,SaksS02},
while it is known that the
space complexity for $0<p\le 2$ is only $\Theta(\eps^{-2}\log(mM) +
\log\log(n))$ bits to achieve success probability
$2/3$ \cite{AMS99, KNW10b}, which can be amplified by outputting the
median estimate of independent
repetitions. In this work,
we focus on this ``feasible'' regime for $p$, $0 < p \le 2$, where 
logarithmic space is achievable.

While
there has been much previous work on minimizing the space consumption
in streaming algorithms, only recently have researchers begun to work
toward minimizing {\em update time} \cite[Question 1]{IITK}, i.e.,
the
time taken to process a new update in the stream. We argue however
that update time itself is an important parameter to optimize, and in
some scenarios it may even be desirable to sacrifice space for speed.
For example, in
network traffic monitoring applications each packet is an update, and
thus it is important that a streaming algorithm processing the packet
stream be able to operate at network speeds (see for example the
applications in \cite{KSZC03,ThorupZhang04}).
Note that if an algorithm has update time say,
$\Omega(1/\eps^2)$,
then achieving a small error parameter such as $\eps=.01$ could be
intractable since this time is multiplied by the length of the stream. This
is true even if the space required of the algorithm is small enough to fit in
memory.

For
$p=2$, it is known that optimal space and $O(1)$ update time are
simultaneously achievable \cite{CCF02, ThorupZhang04}, improving
upon the original $F_2$ algorithm of Alon, Matias, and Szegedy
\cite{AMS99}. For $p=1$
it is known that near-optimal, but not quite optimal, space and
$O(\log(n/\eps))$ update time
are achievable \cite{NW10}.  Meanwhile, optimal (or even near-optimal)
space for other
$p\in (0,2]$ is only known to be achievable with $\poly(1/\eps)$
update time \cite{KNW10b}.

\begin{center}
\begin{figure}
\begin{center}
\begin{small}
\begin{tabular}{|lllll|}
\hline
Paper & Space & Update Time & Model & Which $p$
\\ \hline \cite{AMS99} & $O(\eps^{-2}\log(mM))$ &
$O(\eps^{-2})$ & unrestricted updates & $p = 2$
\\ \hline \cite{CCF02,ThorupZhang04} & $O(\eps^{-2}\log(mM))$ &
$O(1)$ & unrestricted updates & $p = 2$
\\ \hline \cite{FKSV02} & $O(\eps^{-2}\log(mM))$ & $O(\eps^{-2})$ & $\le 2$ updates per coordinate & $p = 1$
\\ \hline \cite{Indyk06,Li08b} & $O(\eps^{-2} \log (n) \log (mM))$ & $O(\eps^{-2})$ & unrestricted updates & $p \in (0,2)$
\\ \hline \cite{KNW10b} & $O(\eps^{-2} \log (mM))$ & $\tO(\eps^{-2})$& unrestricted updates & $p \in (0,2)$
\\ \hline \cite{nw09} & $O(\eps^{-2}\log(mM)\log(1/\eps))$ & $O(\log^2(mM))$ & $\le 2$ updates per coordinate & $p = 1$
\\ \hline \cite{NW10} & $O(\eps^{-2}\log(mM)\log(n))$ & $O(\log(n/\eps))$ & unrestricted updates & $p = 1$
\\ \hline this work & $O(\eps^{-2}\log(mM))$ & $\tO(\log^2(1/\eps))$ &
unrestricted updates & $p \in (0,2)$\\
\hline
\end{tabular}
\end{small}
\caption{Comparison of our contribution to previous works
  on $F_p$-estimation in data streams. All space bounds hide an
  additive $O(\log\log n)$ term.}\FigureName{prev-work}
\end{center}
\end{figure}
\end{center}

\noindent \textbf{Our Contribution: } 
For all $0< p < 2$ and $0<\eps<1/2$ we give an algorithm for
$(1\pm\eps)$-approximating $F_p$ with success probability at least
$2/3$ which uses an optimal
$O(\eps^{-2}\log(mM) + \log\log n)$
bits of space
with $O(\log^2(1/\eps)\log\log(1/\eps))$ update time.\footnote{Throughout this
  document we
    say $g = \tilde{O}(f)$ if $g = O(f\cdot
    \mathrm{polylog}(f))$. Similarly, $g = \tilde{\Omega}(f)$ if $g =
    \Omega(f / \mathrm{polylog}(f))$.} This is a nearly
exponential
improvement in the time complexity of the previous 
space-optimal algorithm for every such $p$.
\vspace{.1in}

\subsection{Previous Work}\SectionName{related}
The complexity of streaming algorithms for moment estimation has a long
history; see \Figure{prev-work} for a comparison of our result
to that of previous work. 

Alon, Matias, and Szegedy were the first to study moment estimation in
data streams \cite{AMS99} and gave a space-optimal algorithm for
$p=2$.  The update
time was later brought down to an optimal $O(1)$ implicitly in \cite{CCF02} and
explicitly in \cite{ThorupZhang04}.  The work of \cite{FKSV02} gave a
space-optimal algorithm for $p=1$, but under the restriction that each
coordinate is updated at most twice, once positively and once
negatively. Indyk \cite{Indyk06} later removed this restriction, and
also gave an algorithm handling all $0<p<2$, but
at the expense of increasing the space by a $\log n$ factor. Li later
\cite{Li08b} provided alternative estimators for all $0 < p < 2$, based on
Indyk's sketches. The
extra $\log n$ factor in the space of these algorithms was later removed 
in \cite{KNW10b}, yielding optimal space.  The
algorithms of \cite{FKSV02,Indyk06, KNW10b, Li08b} all required $\poly(1/\eps)$
update time. 
Nelson and Woodruff \cite{nw09} gave an algorithm for $p=1$ in
the restricted setting where each coordinate is updated at most twice,
as in \cite{FKSV02}, with space suboptimal by a $\log(1/\eps)$ factor,
and with update time $\log^2(mM)$.
They also later gave an algorithm for
$p=1$ with unrestricted updates which was suboptimal by a $\log n$
factor, but had update time
only $O(\log (n/\eps))$ \cite{NW10}.  

On the lower bound front, a lower bound of
$\Omega(\min\{n,m,\eps^{-2}\log(\eps^2mM)\} +
\log\log (nmM))$ was shown in \cite{KNW10b}, together with an upper
bound of $O(\eps^{-2}\log(mM) + \log\log n)$ bits. For nearly the full
range
of parameters these are tight, since if $\eps \le 1/\sqrt{m}$ we can 
store the entire stream in memory in $O(m\log(nM)) =
O(\eps^{-2}\log(nM))$ bits of space (and we can ensure $n = O(m^2)$
via FKS hashing \cite{FredmanKS84} with just an additive $O(\log\log
n)$ bits increase in space), and if $\eps \le 1/\sqrt{n}$ we
can store the entire vector in memory in $O(n\log(mM)) =
O(\eps^{-2}\log(mM))$ bits. Thus, a gap exists only when $\eps$ is
very near $1/\sqrt{\min\{n,m\}}$.
 This lower bound followed many previous lower bounds for this
 problem,
given in \cite{AMS99,BarYossefThesis,JayramKS08,W04,WoodruffThesis}.
For the case $p>2$ it was shown that $\Omega(n^{1-2/p})$ space is
required \cite{BJKS04,cks03,Gronemeier09,Jayram09,SaksS02}, and this was
shown to be tight up to $\poly(\log(nmM)/\eps)$ factors
\cite{bgks06,IW05}.

\subsection{Overview of our approach}
At the top level, our algorithm follows the general approach set
forth by \cite{NW10} for 
$F_1$-estimation.  In that work, the coordinates $i \in
\{1,\ldots,n\}$ were split up into {\em heavy
  hitters}, and the remaining {\em light} coordinates.
A {\em $\phi$-heavy
  hitter} with respect to $F_p$ is a coordinate $i$ such that $|x_i|^p
\ge \phi\|x\|_p^p$. A list $L$ of $\eps^2$-heavy hitters with respect
to $F_1$ were found by running the $\countmin$ sketch of \cite{CM05}.

To estimate the contribution of the light elements to $F_1$,
\cite{NW10} used $R = \Theta(1/\eps^2)$
 independent Cauchy sketches $D_1,\ldots,D_R$ (actually,
$D_j$ was a tuple of $3$ independent Cauchy sketches).
A {\em Cauchy sketch} of a vector $x$, introduced by Indyk
\cite{Indyk06}, is the dot product of $x$ with a random vector $z$
with
independent entries distributed according to the Cauchy distribution.
This distribution has the property that $\inprod{z,x}$ is itself a
Cauchy random variable, scaled by $\|x\|_1$.
Upon receiving an update to $x_i$
in the stream, the update was fed to $D_{h(i)}$ for some hash function
$h:[n]\rightarrow [R]$. At the end of the stream, the estimate of the
contribution to $F_1$ from light elements was $(R / (R - |h(L)|))\cdot
\sum_{j\notin h(L)} \mathsf{EstLi}_1(D_j)$, where $\mathsf{EstLi}_p$
is Li's geometric mean estimator for $F_p$ \cite{Li08b}.
The analysis of \cite{NW10} only used that Li's
geometric mean estimator is unbiased and has a good variance bound.

Our algorithm $\lightest$
for estimating the contribution to $F_p$ from light coordinates
for $p\neq 1$ follows the same approach.  Our main contribution here is
to show that a variant of Li's geometric mean estimator has bounded
variance and is approximately unbiased (to within relative error
$\eps$) even when the associated
$p$-stable random variables are only $k$-wise independent for $k =
\Omega(1/\eps^p)$. This variant allows us to avoid Nisan's pseudorandom
generator \cite{Nisan92} and thus achieve optimal space. While
the work of \cite{KNW10b} also provided an estimator avoiding
Nisan's pseudorandom generator, their estimator is not known to be 
approximately unbiased, which makes it less useful in applications
involving the average of many such estimators. We evaluate
the necessary $k$-wise independent hash function quickly by a combination of
buffering and fast multipoint evaluation of a collection of
pairwise independent polynomials.  Our proof that bounded independence
suffices uses the FT-mollification approach introduced in
\cite{KNW10b} and refined in \cite{DKN10}, which is a method for
showing that the expectation of some function is approximately
preserved by bounded independence, via a smoothing operation
(FT-mollification) and
Taylor's theorem. One
novelty is that while \cite{DKN10,KNW10b} only ever dealt with
FT-mollifying indicator functions of regions in Euclidean space, here
we must FT-mollify functions of the form $f(x) =
|x|^{1/t}$ . To
achieve our results, we express $\E[f(x)] = \int_0^\infty
f(x)\varphi_p(x)dx$ as $\int_{0}^{\infty} f'(x)(1 - \Phi_p(x))dx$
via integration by parts, where $\varphi_p$ is the density function of
the absolute value of the $p$-stable distribution, and $\Phi_p$ is the
corresponding cumulative distribution function. We then note $1 -
\Phi_p(x)
= \Pr[|X| \ge x] = \E[I_{[x,\infty)\cup (-\infty,-x]}(X)]$ for $X$
$p$-stable, where $I_S$ is the indicator function of the set $S$.
We then FT-mollify
$I_{[x,\infty)\cup (-\infty,-x]}$, which {\em is} the indicator
function of some set, to
write $\E[f(x)]$ as a
weighted integral of indicator functions, from which point we can apply
the methods of \cite{DKN10,KNW10b}.

In order to estimate the contribution to $F_p$ from coordinates in
$L$, we
develop a novel data structure we refer to as $\highend$. 
Suppose $L$ contains all the $\alpha$-heavy hitters, and every index
in $L$ is an $(\alpha/2)$-heavy hitter. We would like to compute
$\|x_L\|_p^p \pm O(\eps)\cdot \|x\|_p^p$, where $\alpha =
\Omega(\eps^2)$.
We maintain
a matrix of counters $D_{j,k}$ for $(j,k)\in [t] \times [s]$ for $t =
O(\log(1/\eps))$ and $s = O(1/\alpha)$. For each $j\in[t]$ we have a
hash function $h^j:[n]\rightarrow [s]$ and $g^j:[n]\rightarrow
[r]$ for $r = O(\log(1/\eps))$. The counter $D_{j,k}$ then stores
$\sum_{h^j(v) = k} e^{2\pi i g^j(v)/r} x_v$ for $i =
\sqrt{-1}$.  That is, our data structure is similar to the
$\countsketch$ data structure of Charikar, Chen, and Farach-Colton
\cite{CCF02}, but rather than taking the dot product with a random
sign vector in each counter,  we take the dot product with a vector
whose entries are random
complex roots of unity. At the end of the stream, our estimate of
the $F_p$-contribution from heavy hitters is
$$
\Real \left [\sum_{w\in L}\left(\frac 3t\sum_{k=1}^{t/3}
    e^{-2 \pi i
      g^{j(w,k)}(w) /r}
    \cdot \sign(x_w) \cdot D_{j(w,k), h^{j(w,k)}(w)} \right )^p \right ]
.$$
The choice to use complex roots of unity is to ensure that our estimator 
is approximately unbiased, stemming from the fact that the real
part of large powers of roots of unity is still $0$ in expectation.
Here ${\bf Re}[z]$ denotes the real part of $z$, and $j(w,k)$ denotes
the $k$th smallest value $b\in [t]$ such that $h^b$ isolates $w$ from
the other $w'\in L$ (if fewer than $t/3$ such $b$ exist, we fail). The
subroutine $\filter$ for estimating the heavy hitter contribution for
$p=1$ in
\cite{NW10} did not use complex random variables, but rather just used
the dot product with a random sign vector as in
$\countsketch$. Furthermore, it required a $O(\log(1/\eps))$ factor
more space even for $p=1$, since it did not average estimates across
$\Omega(t)$ levels to reduce variance.

For related problems, e.g., estimating $F_p$ for $p > 2$, using 
complex roots of unity leads
to sub-optimal bounds \cite{g04}. Moreover, it seems that ``similar'' 
algorithms using sign variables in place of roots of unity 
do not work, as they have a constant factor bias in their expectation
for which it is unclear how to remove.
Our initial intuition was that an algorithm
using $p$-stable random variables would be necessary to estimate the
contribution to $F_p$ from the heavy hitters. However, such approaches
we explored suffered from too large a variance.

In parallel we must run an algorithm we develop to {\em find} the
heavy hitters. Unfortunately, this algorithm, as well as $\highend$,
use suboptimal space.
To overcome this, we actually use
a list of $\epsilon^2$-heavy hitters for $\epsilon =
\eps\cdot \log(1/\eps)$.  This then improves the space, at the
expense of increasing the variance of $\lightest$.
We then run $O((\epsilon/\eps)^2)$ pairwise independent instantiations of
$\lightest$ in
parallel and take the average estimate, to bring the variance down.
This increases some part of the update time of $\lightest$ by a
$\log^2(1/\eps)$ factor, but this term turns out to anyway be
dominated by the time to evaluate various hash functions.
Though, even in the extreme case of balancing with $\epsilon =
1$, our algorithm for finding the heavy hitters algorithm requires
$\Omega(\log(n)\log(mM))$ space,
which is suboptimal.  We remedy this by performing a dimensionality
reduction down to dimension $\poly(1/\eps)$ via hashing and dot
products with random
sign vectors.  We then apply $\highend$ to estimate the contribution
from heavy hitters in this new vector, and we show that
with high probability the correctness of our overall algorithm is
still maintained.

\subsection{Notation}\SectionName{notation}
For a positive integer $r$, we use $[r]$ to denote the set
$\{1,\ldots,r\}$.  All logarithms are base-$2$ unless otherwise
noted. For a complex number $z$, $\Real[z]$ is the real
part of $z$, $\Imag[z]$ is the imaginary part of $z$, $\bar{z}$ is the
complex conjugate of $z$, and $|z| \eqdef \sqrt{\bar{z}z}$. 
At times we consider random variables $X$ taking on {\em complex}
values.  For such random variables, we use $\Var[X]$ to denote
$\E[|X - \E[X]|^2]$. Note that the usual statement of Chebyshev's
inequality
still holds under this definition. 

For $x\in
\R^n$ and $S\subseteq [n]$, $x_S$ denotes the $n$-dimensional vector
whose $i$th coordinate is $x_i$ for $i\in S$ and $0$
otherwise. For a probabilistic
event $\mathcal{E}$, we use $\indicate_{\mathcal{E}}$ to denote the
indicator random variable for $\mathcal{E}$. We sometimes refer to a
constant as {\em universal} if it does not depend on other parameters,
such as $n,m,\eps$, etc.
All space bounds are measured in bits.  When measuring time
complexity, we assume a word RAM with machine word size
$\Omega(\log(nmM))$ so that standard arithmetic and bitwise operations
can be performed on words in constant time.
We use {\em reporting time} to refer to the time taken for a
streaming algorithm to answer some query (e.g., ``output an estimate
of $F_p$''). 

Also, we can assume $n =
O(m^2)$  by FKS hashing  \cite{FredmanKS84} with an additive
$O(\log\log n)$ term in our final space bound; see
  Section A.1.1 of \cite{KNW10b} for details.  Thus, henceforth any
  terms involving $n$ appearing in space and time bounds may be
  assumed at most $m^2$.
We also often assume that $n$, $m$, $M$, $\eps$, and $\delta$ are
powers of $2$ (or sometimes $4$), and that $1/\sqrt{n} < \eps <
\eps_0$ for some
universal constant $\eps_0>0$. These assumptions are
without loss of generality. We can assume $\eps > 1/\sqrt{n}$ since
otherwise we could store $x$ explicitly in memory using
$O(n\log(mM)) = O(\eps^{-2}\log(mM))$ bits with constant update and
reporting times. Finally, we assume
$\|x\|_p^p \ge 1$.  This is because, since $x$ has integer entries,
either $\|x\|_p^p \ge 1$, or it is $0$.  The case that it is $0$ only
occurs when $x$ is the $0$ vector, which can be detected in
$O(\log(nmM))$ space by the AMS sketch \cite{AMS99}.

\subsection{Organization}\SectionName{organization}
An ``$F_p$ $\phi$-heavy hitter'' is an index $j$ such that $|\x_j| \ge
\phi\|x\|^p_p$. Sometimes we drop the ``$F_p$'' if $p$ is understood
from context.
In \Section{heavy-contrib}, we give an efficient subroutine $\highend$
for estimating $\|x_L\|_p^p$ to within additive error $\eps\|x\|_p^p$,
where  $L$ is a list containing all $\alpha$-heavy hitters for some
$\alpha>0$, with the promise that no
$i\in L$ is not an $\alpha/2$-heavy hitter.
In \Section{light-contrib} we give a subroutine $\lightest$ for
estimating $\|x_{[n]\backslash L}\|_p^p$.  Finally,
in \Section{final-alg}, we put everything together in a way that
achieves optimal space and fast update time. We discuss how to compute
$L$ in \Section{fp-hh}.

\section{Estimating the contribution from heavy hitters}\SectionName{heavy-contrib}
Before giving our algorithm $\highend$ for estimating $\|x_L\|_p^p$,
we first give a few necessary lemmas and theorems.

The following theorem gives an algorithm for finding the $\phi$-heavy
hitters with respect to $F_p$.
This algorithm uses the dyadic interval idea of \cite{CM05} together
with a black-box reduction of the problem of finding $F_p$ heavy
hitters to the problem of estimating $F_p$.
Our proof is in \Section{fp-hh}. We note that our data structure both
improves
and generalizes that of \cite{GSS08}, which gave an algorithm with
slightly worse bounds that only worked in the case $p=1$.

\begin{theorem}\TheoremName{fp-hh}
There is an algorithm $\fphh$ satisfying the following properties.
Given $0<\phi<1$ and $0<\delta<1$, with probability
at least $1-\delta$,
$\fphh$ produces a list $L$ such that $L$
contains all $\phi$-heavy hitters and does not contain indices which
are not $\phi/2$-heavy hitters. For each $i\in L$, the algorithm also
outputs $\sign(x_i)$, as well as an estimate $\tilde{x}_i$ of $x_i$
satisfying $\tilde{x}_i^p \in [(6/7)|x_i|^p, (9/7)|x_i|^p]$.
Its space usage is $O(\phi^{-1}\log (\phi n)\log(nmM)\log(\log
(\phi n)/(\delta\phi)))$. Its update time is $O(\log
(\phi n)\cdot \log(\log (\phi n)/(\delta\phi))$. Its
reporting time is 
$O(\phi^{-1}(\log (\phi n)\cdot \log(\log (\phi n)/(\delta\phi))))$.
\end{theorem}

The following moment bound can be derived from the Chernoff bound via
integration, and is most likely standard though we do not know the
earliest reference. A proof can be found in 
\cite{KN10}.

\begin{lemma}\LemmaName{good-moment}
Let $X_1,\ldots,X_n$ be such that $X_i$ has expectation $\mu_i$ and
variance $\sigma_i^2$, and $X_i \le K$ almost surely.  Then if the
$X_i$ are $\ell$-wise independent for some even integer $\ell\ge 2$,
$$ \E\left[\left(\sum_{i=1}^n X_i - \mu\right)^\ell\right] \le
2^{O(\ell)} \cdot  \left(\left(\sigma\sqrt{\ell}\right)^\ell  +
  \left(K\ell\right)^\ell\right) ,$$
where $\mu = \sum_i \mu_i$ and $\sigma^2 = \sum_i \sigma_i^2$. In
particular,
$$\Pr\left[\left|\sum_{i=1}^n X_i - \mu\right| \ge \lambda\right] \le
2^{O(\ell)} \cdot  \left(\left(\sigma\sqrt{\ell}/\lambda\right)^\ell  +
  \left(K\ell/\lambda\right)^\ell\right) ,
$$
by Markov's inequality on the random variable $(\sum_i X_i - \mu)^\ell$.
\end{lemma}

\begin{lemma}[Khintchine inequality
  {\cite{Haagerup82}}]\LemmaName{khintchine}
For $x\in \R^n$, $t\ge 2$, and uniformly random
$z\in\{-1,1\}^n$, $\E_z[|\inprod{x,z}|^t] \le \|x\|_2^t
\cdot\sqrt{t}^t$.
\end{lemma}

In the following lemma, and henceforth in this section, $i$ denotes
$\sqrt{-1}$.

\begin{lemma}\LemmaName{unity}
Let $x\in\R^n$ be arbitrary. Let 
$z \in \{e^{2 \pi i/r}, e^{2\pi i \cdot 2/r}, e^{2 \pi i \cdot 3/r},
\ldots, e^{2 \pi i \cdot r/r}\}^n$
be a random such vector for $r\ge 2$ an even integer.
Then for $t\ge 2$ an even integer, $\E_z[|\inprod{x,z}|^t] \le
\|x\|_2^t\cdot 2^{t/2}\sqrt{t}^t$.
\end{lemma}
\begin{proof}
Since $x$ is real, $\left | \langle x, z \rangle \right |^2 = 
\left (\sum_{j=1}^n {\bf Re}[z_j] \cdot x_j \right )^2
+ \left (\sum_{j = 1}^n {\bf Im}[z_j] \cdot x_j \right )^2.$
Then by Minkowski's inequality, 
\begin{align}
\nonumber \E[|\inprod{x,z}|^t] &= \E\left[\left|\left
  (\sum_{j=1}^n {\bf Re}[z_j] \cdot x_j \right )^2
+ \left (\sum_{j = 1}^n {\bf Im}[z_j] \cdot x_j \right
)^2\right|^{t/2}\right] \\
\nonumber & \le \left(2\cdot \max\left\{\E\left[\left(\sum_{j=1}^n
  {\bf Re}[z_j]
\cdot x_j \right)^t \right]^{2/t}, \E\left[\left(\sum_{j=1}^n
  {\bf Im}[z_j]
\cdot x_j \right)^t \right]^{2/t}\right\}\right)^{t/2}\\
& \le 2^{t/2}\cdot \left(\E\left[\left(\sum_{j=1}^n
  {\bf Re}[z_j]
\cdot x_j \right)^t \right]
+ \E\left[\left(\sum_{j = 1}^n {\bf Im}[z_j] \cdot x_j 
\right)^t\right] \right).\EquationName{khintchine-it}
\end{align}
Since $r$ is even, we may write ${\bf Re}[z_j]$ as $(-1)^{y_j}|{\bf Re}[z_j]|$ and
${\bf Im}[z_j] $ as $(-1)^{y_j'}|{\bf Im}[z_j]|$, where $y, y' \in \{-1,1\}^n$
are random sign vectors 
chosen independently of each other. Let us fix the values of
$|{\bf Re}[z_j]|$
and $|{\bf Im}[z_j]|$ for each $j \in [n]$, considering just the
randomness of $y$ and $y'$. Applying  
\Lemma{khintchine} to bound each of the expectations in
\Equation{khintchine-it}, we obtain the bound
$2^{t/2}\cdot\sqrt{t}^t\cdot (\|b\|_2^t + \|b'\|_2^t) \le
2^{t/2}\cdot\sqrt{t}^t\cdot (\|b\|_2^2 + \|b'\|_2^2)^{t/2}$ where $b_j
= \Real[z_j]\cdot x_j$ and $b'_j = \Imag[z_j]\cdot x_j$.  But this is
just $2^{t/2}\cdot\sqrt{t}^t\cdot \|x\|_2^t$ since $|z_j|^2 = 1$.
\end{proof}

\subsection{The $\highend$ data structure}\SectionName{head-contrib}
In this section, we assume we know a subset $L \subseteq [n]$ of indices $j$ so
that 
\begin{enumerate}
\item for all $j$ for which $|\x_j|^p \geq \alpha \|\x\|^p_p$, $j \in L$, 
\item if $j \in L$, then $|\x_j|^p \geq (\alpha/2) \|\x\|^p_p$,
\item for each $j \in L$, we know $\sign(x_j)$. 
\end{enumerate}
for some $0<\alpha < 1/2$ which we know.  We also are given some $0 <
\eps < 1/2$.
We would like to output a value $\|x_L\|_p^p \pm O(\eps)\|x\|_p^p$
with large constant probability. We assume $1/\alpha = O(1/\eps^2)$.

We first define the $\basichighend$ data structure. 
Put $s = \ceil{4/\alpha}$.
We choose a hash function $h:[n] \rightarrow [s]$ at random from an
$r_h$-wise independent family for $r_h = \Theta(\log(1/\alpha))$.
Also, let $r = \Theta(\log 1/\eps)$ be a sufficiently large even
integer.
For each $j \in [n]$, we associate a random complex root of unity 
$e^{2 \pi i g(j) /r}$, where $g:[n]\rightarrow[r]$ is drawn at random
from an $r_g$-wise independent family for $r_g = r$.
We initialize $s$
counters $b_1, \ldots, b_s$ to $0$. Given an update of the form
$(j,v)$, add $e^{2 \pi i g(j) /r} \cdot v$ to $b_{h(j)}$.

We now define the $\highend$ data structure as follows. 
Define $T = \tau\cdot \max\{\log(1/\eps),\log(2/\alpha)\}$ for a
sufficiently large constant $\tau$ to be determined
later. Define
$t = 3T$ and instantiate $t$ independent copies of the 
$\basichighend$ data structure. 
Given an update $(j,v)$, perform
the update described above to each of the copies of $\basichighend$.
We think of this data structure as a $t \times s$ matrix of counters
$D_{j,k}$, $j \in [t]$ and $k \in [s]$. We let $g^j$ be the hash function
$g$ in the $j$th independent instantiation of $\basichighend$, and
similarly define $h^j$. We sometimes use $g$ to denote the tuple
$(g^1,\ldots,g^t)$, and similarly for $h$.

We now define our estimator, but first we give some notation.  For
$w\in L$, let $j(w,1)<j(w,2)<\ldots<j(w,n_w)$ be the set of $n_w$
indices $j\in
[t]$ such that $w$ is {\em isolated} by $h^j$ from other indices in
$L$; that is, indices $j\in [t]$ where no other $w'\in L$ collides
with $w$ under $h^j$.   

\vspace{.1in}

\noindent {\bf Event $\mathcal{E}$}. Define $\mathcal{E}$ to be the
event that $n_w \ge T$ for all $w\in L$.

\vspace{.1in}

If $\mathcal{E}$ does not hold, our estimator simply fails.  Otherwise, 
define
$$ x_w^* = \frac 1T\cdot \sum_{k=1}^T e^{-2 \pi i
      g^{j(w,k)}(w) /r} \cdot \textrm{sign}(x_w) \cdot D_{j(w,k),
      h^{j(w,k)}(w)} .$$
If $\Real [x_w^*] < 0$ for any $w\in L$, then we output
fail. Otherwise, define
$$ \Psi' = \sum_{w \in L} \left( x_w^*
  \right )^p
.$$
Our estimator is then $\Psi = \Real[\Psi']$.
Note $x^*$ is a complex number.  By $z^p$ for complex $z$, we mean
$|z|^p \cdot e^{ip\cdot \arg(z)}$, where 
$\arg(z)\in (-\pi, \pi]$ is the angle formed by the vector from the
origin to $z$ in the complex plane.

\subsection{A useful random variable}
For $w\in L$, we make the definitions
$$y_w \eqdef \frac{x_w^* - |x_w|}{|x_w|}, \hspace{.5in} \Phi_w \eqdef
|x_w|^p\cdot
\left(\sum_{k=0}^{r/3}\binom{p}{k}\cdot y_w^k \right) $$
as well as $\Phi \eqdef \sum_{w\in L} \Phi_w$.
We assume $\mathcal{E}$ occurs so that the $y_w$ and $\Phi_w$
(and hence $\Phi$) are defined. Also, we use the definition
$\binom{p}{k} = (\prod_{j=0}^{k-1} (p - j))/k!$ (note $p$ may not be
an integer).

Our overall goal is to show that $\Psi = \|x_L\|_p^p \pm O(\eps)\cdot
\|x\|_p^p$ with large constant probability.  Our proof plan is
to first show that $|\Phi - \|x_L\|_p^p| = O(\eps)\cdot \|x\|_p^p$
with large constant probability, then
to show that $|\Psi' - \Phi| = O(\eps)\cdot \|x\|_p^p$ with large
constant probability, at which point our claim follows by a union
bound and the triangle inequality since $|\Psi - \|x_L\|_p^p| \le
|\Psi' - \|x_L\|_p^p|$ since $\|x_L\|_p^p$ is real.

Before analyzing $\Phi$, we define the following event. 

\vspace{.1in}

\noindent {\bf Event $\mathcal{D}$}. Let $\mathcal{D}$ be the event
that for all $w\in L$ we have
$$\frac{1}{T^2} \sum_{k=1}^T \sum_{\substack{v\notin
    L\\h^{j(w,k)}(v) = h^{j(w,k)}(w)}} x_v^2 <
\frac{(\alpha \cdot \|x\|_p^p)^{2/p}}{r} .$$

\vspace{.1in}

We also define
$$V =
\frac{1}{T^2}\sum_{w\in L}\sum_{j=1}^t\sum_{\substack{v\notin
    L\\h^j(w) = h^j(v)}} |x_w|^{2p-2}\cdot |x_v|^2 .$$

\begin{theorem}\TheoremName{phiworks}
Conditioned on $h$,
$\E_g[\Phi] = \|\x_L\|_p^p$ and $\Var_g[\Phi\mid \mathcal{D}] = O(V)$.
\end{theorem}
\begin{proof}
By linearity of expectation,
\begin{eqnarray*}
\E_g[\Phi] = \sum_{w \in L} |x_w|^p \cdot 
\left [\sum_{k=0}^{r/3} {p \choose k} \E_g[y_w^k]\right]
= \sum_{w \in L} |x_w|^p + 
\sum_{w \in L} |x_w|^p \cdot \sum_{k = 1}^{r/3} {p \choose r}
\E_g\left [y_w^k \right ] ,
\end{eqnarray*}
where we use that ${p \choose 0} = 1$. Then
$\E_g[y_w^k] = 0$ for $k > 0$ by 
using linearity of expectation and $r_g$-wise independence, 
since each summand involves at most
$k < r$ $r$th roots of unity. 
Hence,
\begin{eqnarray*}
\E_g[\Phi] = \sum_{w \in L}|x_w|^p.
\end{eqnarray*}
We now compute the variance. Note that if the $g^j$ were each fully
independent, then we would have $\Var_g[\Phi\mid \mathcal{D}] =
\sum_{w\in
  L}\Var_g[\Phi_w\mid \mathcal{D}]$ since different $\Phi_w$ depend on
evaluations of the $g^j$ on
disjoint $v\in [n]$.  However, since $r_g > 2r/3$,
$\E_g[|\Phi|^2]$
is identical as in the case of full independence of the $g^j$. We thus
have $\Var_g[\Phi \mid\mathcal{D}] = \sum_{w\in L}
\Var_g[\Phi_w\mid\mathcal{D}]$ and have reduced to
computing $\Var_g[\Phi_w \mid \mathcal{D}]$.
\begin{eqnarray*}
\Var_g[\Phi_w\mid\mathcal{D}] & = & \E_g[|\Phi_w  -
\E_g[\Phi_w]|^2 \mid\mathcal{D}]\\
&=& |x_w|^{2p}\cdot \E_g\left[\left|\sum_{k=1}^{r/3} \binom{p}{k}
    y_w^k\right|^2\mid \mathcal{D}\right]\\
&=& |x_w|^{2p}\cdot \left(p^2\cdot\E_g[|y_w|^2\mid \mathcal{D}] +
  \sum_{k=2}^{r/3}
O(\E_g[|y_w|^{2k} \mid \mathcal{D}])\right)
\end{eqnarray*}
We have
\begin{equation}
 \E_g[|y_w|^2 \mid\mathcal{D}] 
\eqdef u_w^2
= \frac {1}{T^2}\sum_{k=1}^T
\sum_{\substack{v\notin
    L\\h^{j(w,k)}(v) = h^{j(w,k)}(w)}} \frac{x_v^2}{x_w^2}
,\EquationName{uw}
\end{equation}
so that
$$ \sum_{w\in L} p^2\cdot \E_g[|y_w|^2 \mid \mathcal{D}] \le p^2V .$$
\Equation{uw} follows since, conditioned on $\mathcal{E}$ so that
$y_w$ is well-defined,
$$ \E_g[|y_w|^2] =
\frac{1}{T^2x_w^2}\sum_{k=1}^T\sum_{k'=1}^T\sum_{\substack{v\notin
    L\\h^{j(w,k)}(v) = h^{j(w,k)}(w)}}\sum_{\substack{v'\notin
    L\\h^{j(w,k')}(v') = h^{j(w,k')}(w)}}\E[e^{-2\pi i(g^{j(w,k)}(v) -
  g^{j(w,k')}(v'))/r}]x_vx_{v'} .$$
When $j(w,k)\neq j(w,k')$ the above expectation is $0$ since the
$g^j$ are independent across different $j$. When $j(w,k) = j(w,k')$
the above expectation is only non-zero for $v= v'$ since 
$r_g \ge 2$.

We also have for $k\ge 2$ that
$$ \E_g[|y_w|^{2k} \mid \mathcal{D}] \le 2^{O(k)}\cdot u_w^{2k}\cdot
(2k)^k$$
by \Lemma{unity}, so that
$$ \sum_{k=2}^{r/3} \E_g[|y_w|^{2k}\mid \mathcal{D}] = O(u_w^2) $$
since $\mathcal{D}$ holds and so the sum is dominated by its first
term. Thus, $\Var_g[\Phi \mid \mathcal{D}] =
O(V)$.
\end{proof}

\begin{lemma}\LemmaName{Visgood} $\E_h[V] \le 3\alpha\cdot
  \|x\|_p^{2p}/(4T)$.
\end{lemma}
\begin{proof}
For any $w\in L$, $v\notin L$, and $j\in [t]$, we have
$\Pr_h[h^j(w) = h^j(v)] = 1/s \le \alpha/4$ since $r_h\ge 2$. Thus,
\begin{align}
\nonumber \E_h[V] & \le \frac{\alpha}{4T^2}\sum_{\substack{w\in
    L\\v\not\in L\\j\in [t]}} |x_w|^{2p-2}|x_v|^2\\
\nonumber {} & = \frac{3\alpha}{4T} \left(\sum_{w\in L} |x_w|^p
  |x_w|^{p-2}\right)\left(\sum_{v\notin L} |x_v|^2\right)\\
{} & \le \frac{3\alpha}{4T} \left(\sum_{w\in
    L}\|x\|_p^p(\alpha \cdot\|x\|_p^p)^{(p-2)/p}\right)\left(\frac
  1{\alpha}(\alpha
  \cdot\|x\|_p^p)^{2/p}\right)\EquationName{max-l2}\\
\nonumber {} & = \frac 34\cdot \alpha \cdot \|x\|_p^{2p}/T .
\end{align}
where \Equation{max-l2} used that $\|x_{[n]\backslash L}\|_2^2$ is
maximized when $[n]\backslash L$ contains exactly $1/\alpha$
coordinates $v$ each with $|x_v|^p = \alpha\|x\|_p^p$, and that
$|x_w|^{p-2} \le (\alpha\cdot \|x\|_p^p)^{(p-2)/p}$ since $p\le 2$.
\end{proof}

\begin{lemma}\LemmaName{isolated}
$\Pr_h[\mathcal{E}] \ge 1 - \eps$.
\end{lemma}
\begin{proof}
For any $j\in [t]$, the probability that $w$ is isolated by $h^j$ is
at least $1/2$, since the expected number of collisions with $w$ is at
most $1/2$ by pairwise independence of the $h^j$ and the fact that
$|L| \le 2/\alpha$ so that $s\ge 2|L|$. If $X$ is the expected number
of buckets where $w$ is isolated, the Chernoff bound gives $\Pr_h[X <
(1-\epsilon)\E_h[X]] < \exp(-\epsilon^2\E_h[X]/2)$ for $0<\epsilon<1$.  The
claim follows for $\tau \ge 24$ by setting $\epsilon = 1/3$ then
applying a union bound
over $w\in L$.
\end{proof}

\begin{lemma}\LemmaName{condition-D}
$\Pr_h[\mathcal{D}] \ge 63/64$.
\end{lemma}
\begin{proof}
We apply the bound of \Lemma{good-moment} for a single $w\in L$. Define
$X_{j,v} =
(x_v^2/T^2)\cdot \indicate_{h^j(v) = h^j(w)}$ and $X = \sum_{j=1}^t
\sum_{v\notin L} X_{j,v}$. Note that $X$ is an upper bound for the
left hand side of the inequality defining $\mathcal{D}$, and thus it
suffices to show a tail bound for $X$. In the notation of
\Lemma{good-moment}, we have $\sigma^2 \le
(3/(sT^3))\cdot \|x_{[n]\backslash L}\|_4^4$, $K = (\alpha\cdot
\|x\|_p^p)^{2/p}/T^2$, and $\mu =
(3/(sT))\cdot \|x_{[n]\backslash L}\|_2^2$.
Since $\|x_{[n]\backslash L}\|_2^2$ and $\|x_{[n]\backslash L}\|_4^4$
are each maximized when there are exactly $1/\alpha$ coordinates
$v\notin L$ with $|x_v|^p = \alpha\cdot \|x\|_p^p$, 
$$\sigma^2
\le \frac{3}{4T^3}\cdot (\alpha\cdot\|x\|_p^p)^{4/p}, \hspace{.5in}\mu
\le \frac{3}{4T}\cdot (\alpha\cdot \|x\|_p^p)^{2/p} .$$

Setting $\lambda =
(\alpha\cdot \|x\|_p^p)^{2/p}/(2r)$, noting that $\mu <
\lambda$ for $\tau$ sufficiently large, and assuming $\ell\le r_h$ is
even, we apply \Lemma{good-moment} to obtain
$$ \Pr[X \ge 2\lambda] \le 2^{O(\ell)} \cdot
  \left(\left(\frac{\sqrt{3}r\cdot
        \sqrt{\ell}}{T^{3/2}}\right)^\ell +
    \left(\frac{2r\cdot \ell}{T^2}\right)^\ell\right)
  .$$
By setting $\tau$ sufficiently large and $\ell = \log(2/\alpha) + 6$,
the above probability is at most $(1/64)\cdot (\alpha/2)$. The
lemma follows by a union bound over all $w\in L$, since $|L| \le
2/\alpha$.
\end{proof}

We now define another event.

\vspace{.1in}

\noindent {\bf Event $\mathcal{F}$}. Let $\mathcal{F}$ be the event
that for all $w\in L$ we have $|y_w| < 1/2$.

\vspace{.1in}

\begin{lemma}\LemmaName{last-statement}
$\Pr_g[\mathcal{F} \mid \mathcal{D}] \ge 63/64$.
\end{lemma}
\begin{proof}
$\mathcal{D}$ occurring implies that $u_w \le \sqrt{1/r} \le
\sqrt{1/(64(\log(2/\alpha)+6)}$ (recall
we assume $1/\alpha = O(1/\eps^2)$ and pick $r = \Theta(\log(1/\eps))$
sufficiently large, and $u_w$ is as is defined in \Equation{uw}), and
we also have $\E_g[|y_w|^\ell \mid \mathcal{D}]
< u_w^\ell\sqrt{\ell}^\ell 2^\ell$ by \Lemma{unity}.  
Applying Markov's bound on the random variable $|y_w|^\ell$ for even
$\ell \le r_g$, we have $|y_w|^\ell$ is determined by $r_g$-wise
independence of the $g^j$, and thus
$$ \Pr_g[|y_w| \ge 1/2\mid\mathcal{D}] <
\left(\sqrt{\frac{16\ell}{64(\log(2/\alpha)+6)}}\right)^\ell ,$$
which equals $(1/64)\cdot (\alpha/2)$ for $\ell = \log(2/\alpha) +
6$.  We then apply a union bound over all $w\in L$.
\end{proof}

\begin{lemma}\LemmaName{good-estimator}
Given $\mathcal{F}$, $|\Psi' - \Phi| < \eps \|x_L\|_p^p$.
\end{lemma}
\begin{proof}
Observe
$$ \Psi' = \sum_{w\in L}|x_w|^p\cdot (1 + y_w)^p  .$$
We have that $\ln(1+z)$, as a function of $z$, is holomorphic on the
open disk of radius $1$
about $0$ in the complex plane, and thus $f(z) = (1+z)^p$ is
holomorphic in this region
since it is the composition $\exp(p\cdot \ln(1 + z))$ of holomorphic
functions. Therefore, $f(z)$ equals its Taylor expansion about $0$
for all $z\in\C$ with $|z| < 1$ (see for example \cite[Theorem
11.2]{Wong08}). 
Then since $\mathcal{F}$ occurs, we can Taylor-expand 
$f$ about $0$ for $z = y_w$ and
apply Taylor's theorem to obtain
\begin{align*}
 \Psi' & = \sum_{w\in
  L}|x_w|^p\left(\sum_{k=0}^{r/3}\binom{p}{k}y_w^k \pm
O\left(\binom{p}{r/3 + 1}\cdot |y_w|^{-r/3-1}\right)\right) \\
{} & = \Phi + O\left(\|x_L\|_p^p\cdot \left(\binom{p}{r/3 + 1}\cdot
    |y_w|^{-r/3-1}\right)\right)
\end{align*}
The lemma follows since $\binom{p}{r/3+1} < 1$ and $|y_w|^{-r/3-1}
< \eps$ for $|y_w| < 1/2$.
\end{proof}

\begin{theorem}\TheoremName{psiworks}
The space used by $\highend$ is
$O(\alpha^{-1}\log(1/\eps)\log(mM/\eps) + O(\log^2(1/\eps)\log
n))$. The update time is
$O(\log^2(1/\eps))$. The reporting time is
$O(\alpha^{-1}\log(1/\eps)\log(1/\alpha))$. Also, $\Pr_{h,g}[|\Psi -
\|x_L\|_p^p| < O(\eps)\cdot \|x\|_p^p] > 7/8$. 
\end{theorem}
\begin{proof}
We first argue correctness.
By a union bound, $\mathcal{E}$ and $\mathcal{D}$ hold
simultaneously with probability $31/32$.
By Markov's inequality and \Lemma{Visgood},
$V = O(\alpha\cdot \|x\|_p^{2p}/T)$ with probability $63/64$.  We then
have by Chebyshev's inequality and \Theorem{phiworks} that $|\Phi -
\|x_L\|_p^p| = O(\eps)\cdot \|x\|_p^p$ with probability $15/16$.
\Lemma{good-estimator} then implies
$|\Psi' - \|x_L\|_p^p| = O(\eps)\cdot \|x\|_p^p$
with probability $15/16 - \Pr[\neg \mathcal{F}]>
7/8$ by \Lemma{last-statement}. In this case, the same must hold true
for $\Psi$ since $\Psi =
\Real [\Psi']$ and $\|x_L\|_p^p$ is real.

Next we discuss space complexity. We start with analyzing the
precision required to store the counters $D_{j,k}$. Since our
correctness analysis conditions on 
$\mathcal{F}$, we can assume $\mathcal{F}$ holds.
We store the real and imaginary parts of each
counter $D_{j,k}$ separately. If we store each such part to within
precision $\gamma/(2mT)$ for
some $0<\gamma<1$ to be determined later, then each of the real and
imaginary parts, which are
the sums
of at most $m$ summands from the $m$ updates in the stream, is stored
to within additive error $\gamma/(2T)$ at the end of the stream.
Let $\tilde{x}_w^*$ be our calculation of $x_w^*$ with such
limited precision.
Then, each of the real and
imaginary parts of $\tilde{x}_w^*$ is within additive error
$\gamma/2$ of those for $x_w^*$. Since $\mathcal{F}$ occurs, $|x_w^*|
> 1/2$, and thus $\gamma/2 <
\gamma |x_w^*|$, implying $|\tilde{x}_w^*| = (1\pm
O(\gamma))|x_w^*|$.  Now we argue $\arg(\tilde{x}_w^*) =
\arg(x_w^*) \pm O(\sqrt{\gamma})$. Write $x_w^* = a + ib$ and
$\tilde{x}_w^* = \tilde{a} + i\tilde{b}$ with $\tilde{a} = a \pm
\gamma/2$ and $\tilde{b} = b \pm \gamma/2$. 
We have $\cos(\arg(x_w^*)) = a/\sqrt{a^2 + b^2}$.
Also,
$\cos(\arg(\tilde{x}_w^*)) = (a\pm \gamma/2) / ((1\pm
O(\gamma))\sqrt{a^2 + b^2}) = (1\pm O(\gamma))\cos(\arg(x_w^*)) \pm
O(\gamma) = \cos(\arg(x_w^*)) \pm O(\gamma)$, implying
$\arg(\tilde{x}_w^*) = \arg(x_w^*) \pm O(\sqrt{\gamma})$.
Our final output is 
$\sum_{w\in L} |\tilde{x}^*_w|^p\cdot \cos(p\cdot
\arg(\tilde{x}^*_w))$. Since $\cos$ never
has derivative larger than $1$ in magnitude, this is $\sum_{w\in L}
[(1\pm O(\gamma))|x_w^*|^p\cos(p\cdot \arg(x_w^*)) \pm
O(\sqrt{\gamma})\cdot(1\pm O(\gamma))|x_w^*|^p]$. Since $\mathcal{F}$
occurs, $|x_w^*|^p < (3/2)^p\cdot |x_w|^p$, and thus our overall error
introduced from limited precision is $O(\sqrt{\gamma}\cdot
\|x_L\|_p^p)$, and it thus suffices to set $\gamma = O(\eps^2)$,
implying each $D_{j,k}$ requires $O(\log(mM/\eps))$ bits of
precision.
For the remaining part of the space analysis, we discuss storing the
hash functions.
The hash functions $h^j,g^j$ each require $O(\log(1/\eps)\log n)$
bits of seed, and thus in total consume $O(\log^2(1/\eps)\log n)$ bits.

Finally we discuss time complexity. To
perform an
update, for each $j\in [t]$ we must evaluate $g^j$ and $h^j$ then
update a counter.  Each of $g^j,h^j$ require $O(\log(1/\eps))$ time to
evaluate. For the reporting time, we can mark all counters with the
unique $w\in L$ which hashes to it under the corresponding $h^j$ (if a
unique such $w$
exists) in $|L| \cdot t\cdot r_h =
O(\alpha^{-1}\log(1/\eps)\log(1/\alpha))$ time. Then, we sum up the
appropriate counters for each $w\in L$, using the Taylor expansion of
$\cos(p\cdot \arg(z))$ up to the $\Theta(\log(1/\eps))$th degree to
achieve additive error $\eps$.
Note that conditioned on $\mathcal{F}$, $\arg(x_w^*) \in (-\pi/4,
\pi/4)$, so that $|p\cdot \arg(x_w^*)|$ is bounded away from $\pi/2$
for $p$ bounded away from $2$; in fact, one can even show via some
calculus that $\arg(x_w^*)\in (-\pi/6,\pi/6)$ when $\mathcal{F}$
occurs by showing that $\cos(\arg(x_w^*)) = \cos(\arg(1 - y_w))$ is
minimized for $|y_w| \le 1/2$ when $y_w = 1/4 +
i\sqrt{3}/4$. Regardless, additive error $\eps$ is relative error
$O(\eps)$, since if $|p\cdot \arg(z)|$ is bounded away from $\pi/2$,
then $|\cos(p\cdot \arg(z))| = \Omega(1)$.
\end{proof}

\section{Estimating the contribution from light elements}\SectionName{light-contrib}
In this section, we show how to estimate the contribution to $F_p$
from coordinates of $x$ which are not heavy hitters.
More precisely, given a list $L\subseteq[n]$ such that $|L| \le
2/\eps^2$ and $|x_i|^p \le \eps^2 \|x\|_p^p$ for all $i\notin L$, we
describe a subroutine $\lightest$ that outputs a value that is
$\|x_{[n]\backslash L}\|_p^p \pm O(\eps)\cdot \|x\|_p^p$ with
probability at least $7/8$.  This estimator is essentially the same as
that given for $p=1$ in \cite{NW10}, though in this work we show that
(some variant of) the geometric mean estimator of \cite{Li08b}
requires only bounded independence, in order that we may obtain
optimal space.

Our description follows. We first need the following theorem, which
comes from a derandomized variant of the geometric mean estimator.
Our proof is in \Section{gme}.

\begin{theorem}\TheoremName{gme-good}
For any $0 < p < 2$, there is a randomized data structure $D_p$, and a
deterministic
algorithm $\Est_p$ mapping the state space of the data structure to
reals, such that
\begin{enumerate}
\item $\E[\Est_p(D_p(x))] = (1\pm \eps)\|x\|_p^p$
\item $\E[\Est_p(D_p(x))^2] \le C_p\cdot \|x\|_p^{2p}$
\end{enumerate}
for some constant $C_p>0$ depending only on $p$, and where
the expectation is taken over the randomness used by $D_p$. Aside
from storing a length-$O(\eps^{-p}\log(nmM))$
random string, the space complexity is $O(\log(nmM))$. The update time
is the time to evaluate a $\Theta(1/\eps^p)$-wise independent hash
function over a field of size $\poly(nmM)$, and the reporting time is
$O(1)$.
\end{theorem}

We also need the following algorithm for fast multipoint evaluation of
polynomials.

\begin{theorem}[{\cite[Ch. 10]{GG99}}]\TheoremName{fastmult}
Let $\mathbf{R}$ be a ring, and let $q\in \mathbf{R}[x]$ be a
degree-$d$ polynomial. Then, given distinct
$x_1,\ldots,x_d\in\mathbf{R}$, all the values $q(x_1),\ldots,q(x_d)$
can be computed using $O(d\log^2d\log\log d)$ operations
over $\mathbf{R}$.
\end{theorem}

The guarantees of the final $\lightest$ are then given in
\Theorem{modified}, which is a
modified form of an algorithm designed in \cite{NW10} for the case
$p=1$.  A description of the modifications of the algorithm in
\cite{NW10} needed to work for $p\neq 2$ is given in \Remark{modified},
which in part uses the following uniform hash family of Pagh and Pagh
\cite{PP08}.

\begin{theorem}[Pagh and Pagh {\cite[Theorem 1.1]{PP08}}]\TheoremName{pagh}
Let $S \subseteq U = [u]$ be a set of $z>1$ elements, and let $V =
[v]$, with $1<v\le u$. Suppose the machine word size is
$\Omega(\log(u))$.
For any constant $c>0$ there is a word RAM
algorithm that, using
time $\log(z)\log^{O(1)}(v)$ and $O(\log(z) + \log\log(u))$ bits of
space, selects a family $\hash$ of functions from $U$ to $V$
(independent of $S$) such that:
\begin{enumerate}
\item With probability $1 - O(1/z^c)$, $\hash$ is $z$-wise independent
when restricted to $S$.
\item Any $h\in \hash$ can be represented by a RAM data structure
  using $O(z\log(v))$ bits of space, and $h$ can be evaluated in
  constant time after an initialization step taking $O(z)$ time.
\end{enumerate}
\end{theorem}

\begin{theorem}[{\cite{NW10}}]\TheoremName{modified}
Suppose we are given $0<\eps<1$, and given a list $L\subseteq[n]$
at the end of the data stream such that $|L| \le
2/\eps^2$ and $|x_i|^p < \eps^2 \|x\|_p^p$ for all $i\notin L$.
Then, given access to a randomized data structure satisfying
properties (1) and (2) of \Theorem{gme-good},
there is an algorithm $\lightest$ satisfying the
following.
The randomness used by $\lightest$ can be broken up into a certain
random hash function $h$, and another random string $s$.
$\lightest$ outputs a value $\Phi$' satisfying
$\E_{h,s}[\Phi'] = (1\pm O(\eps))\|x_{[n]\backslash
  L}\|_p^p$, and $\E_h[\Var_s[\Phi']] = O(\eps^2 \|x\|_p^{2p})$.
The space usage is $O(\eps^{-2}\log(nmM))$,
the update time is $O(\log^2(1/\eps)\log\log(1/\eps))$, and the
reporting time is
$O(1/\eps^2)$.
\end{theorem}

\begin{remark}\RemarkName{modified}
\textup{
The claim of \Theorem{modified} is not stated in the same form in
\cite{NW10}, and thus we provide some explanation.
The work of
\cite{NW10} only focused on the case $p=1$.
There, in Section 3.2, $\lightest$ was defined\footnote{The estimator
  given there was never actually named, so we name it $\lightest$
  here.} by
creating $R = 4/\eps^2$ independent instantiations of $D_1$,
which we label $D_1^1,\ldots,D_1^R$ ($R$ chosen so that $R \ge 2|L|$),
and
picking a hash function $h:[n]\rightarrow[R]$ from a random hash
family constructed as in \Theorem{pagh} with $z = R$ and $c \ge
2$. Upon receiving an update to
$x_i$ in the stream, the update was fed to $D_1^{h(i)}$. The final
estimate was defined as follows.  Let $I = [R]\backslash h(L)$. Then,
the estimate was $\Phi' = (R/|I|)\cdot \sum_{j\in I} \Est_1(D_1^j)$.
In place of a
generic $D_1$, the presentation in \cite{NW10} used Li's geometric
mean estimator
\cite{Li08b}, though the analysis (Lemmas 7 and 8 of \cite{NW10}) only
made use of the generic
properties of $D_1$ and $\Est_1$ given in \Theorem{gme-good}.
Let $s = (s_1,\ldots,s_R)$ be the tuple of random strings used by the
$D_1^j$, where the entries of $s$ are pairwise independent.
The analysis then showed that (a) $\E_{h,s}[\Phi'] =
(1\pm O(\eps))\|x_{[n]\backslash L}\|_1$, and (b) $\E_h[\Var_s[\Phi']]
= O(\eps^2\|x\|_1^2)$. For (a), the same analysis applies for $p\neq
1$ when using $\Est_p$ and $D_p$ instead.  For (b), it was shown that
$\E_h[\Var_s[\Phi']]
= O(\|x_{[n]\backslash L}\|_2^2 + \eps^2\|x_{[n]\backslash
  L}\|_1^2)$. The same analysis shows that $\E_h[\Var_s[\Phi']]
= O(\|x_{[n]\backslash L}\|_{2p}^{2p} + \eps^2\|x_{[n]\backslash
  L}\|_p^p)$ for $p\neq 1$.  Since $L$ contains all the $\eps^2$-heavy hitters,
$\|x_{[n]\backslash L}\|_{2p}^{2p}$ is maximized when there are
$1/\eps^2$ coordinates $i\in [n]\backslash L$ each with $|x_i|^p =
\eps^2\|x\|_p^p$, in which case $\|x_{[n]\backslash L}\|_{2p}^{2p} =
\eps^2\|x\|_p^{2p}$.
}

\textup{
To achieve the desired update time, we buffer every $d = 1/\eps^p$
updates then perform the fast multipoint evaluation of
\Theorem{fastmult} in batch (note this does not affect our space
bound since $p<2$). That is, although the
hash function $h$ can be evaluated in constant time, updating any
$D_p^j$ requires evaluating a degree-$\Omega(1/\eps^p)$ polynomial,
which na\"{i}vely requires $\Omega(1/\eps^p)$ time.  Note that one
issue is that the different data structures
$D_p^j$ use different polynomials, and thus we may need to evaluate
$1/\eps^p$ different polynomials on the $1/\eps^p$ points, defeating
the purpose of batching.  To remedy this, note that these polynomials
are themselves pairwise independent.  That is, we can assume there are
two coefficient vectors $a,b$ of length $d+1$, and the polynomial
corresponding to $D_p^j$ is given by the coefficient vector $j\cdot a
+ b$.  Thus, we only need to perform fast multipoint evaluation on the
two
polynomials defined by $a$ and $b$. 
To achieve worst-case update time, this computation can
be spread over the next $d$ updates. If a query comes before $d$
updates are batched, we need to perform $O(d\log d\log\log d)$ work at
once, but this is already dominated by our $O(1/\eps^2)$ reporting
time since $p<2$.
}
\end{remark}

\section{The final algorithm: putting it all
  together}\SectionName{final-alg}
To obtain our final algorithm, one option is to run $\highend$ and
$\lightest$ in parallel after finding $L$, then output the sum of
their estimates.  Note that by the variance bound in
\Theorem{modified}, the output of a single instantiation of
$\lightest$ is $\|x_{[n]\backslash L}\|_p^p \pm O(\eps)\|x\|_p^p$ with
large constant probability.  The downside to this option is that
\Theorem{fp-hh} uses space that would make our
overall $F_p$-estimation algorithm suboptimal by $\polylog(n/\eps)$
factors, and $\highend$ by an $O(\log(1/\eps))$ factor for $\alpha =
\eps^2$ (\Theorem{psiworks}).
We can overcome this by a combination of balancing and universe
reduction. Specifically, for balancing,
notice that if instead of having $L$ be a list of $\eps^2$-heavy
hitters, we instead defined it as a list of $\epsilon^2$-heavy hitters for
some $\epsilon > \eps$, we could improve the space of both
\Theorem{fp-hh} and \Theorem{psiworks}.  To
then make the variance in $\lightest$ sufficiently small,
i.e. $O(\eps^2\|x\|_p^2)$, we could run $O((\epsilon / \eps)^2)$
instantiations of $\lightest$ in parallel and output the average
estimate, keeping the space optimal but increasing the update time
to $\Omega((\epsilon/\eps)^2)$. This balancing gives a smooth
tradeoff between space and update time;
in fact note that for $\epsilon = 1$,
our overall algorithm simply becomes a derandomized variant of Li's
geometric mean estimator.  We would like though to have $\epsilon \ll
1$ to have small update time.

Doing this balancing does not
resolve all our issues though, since \Theorem{fp-hh} is
suboptimal by a $\log n$ factor.  That is, even if we picked $\epsilon
= 1$, \Theorem{fp-hh} would cause our overall space to be
$\Omega(\log(n)\log(mM))$, which is suboptimal. To overcome this issue we
use universe reduction.  Specifically, we set $N = 1/\eps^{18}$ and
pick hash functions $h_1:[n]\rightarrow[N]$ and
$\sigma:[n]\rightarrow\{-1,1\}$. We define a new $N$-dimensional
vector $y$ by $y_i = \sum_{h_1(j) = i} \sigma(j) x_j$. Henceforth in
this section, $y$, $h_1$, and $\sigma$ are as discussed here. Rather
than computing a list $L$ of heavy hitters of
$x$,
we instead compute a list $L'$ of heavy hitters of $y$.
Then, since $y$
has length only $\poly(1/\eps)$, \Theorem{fp-hh} is only
suboptimal by $\polylog(1/\eps)$ factors and our balancing trick
applies. The list $L'$ is also used in place of $L$ for both
$\highend$ and $\lightest$.
Though, since we never learn $L$, we
must modify the algorithm $\lightest$ described in \Remark{modified}.
Namely, the hash
function $h:[n]\rightarrow[R]$ in \Remark{modified} should be
implemented as the composition of $h_1$, and a hash function
$h_2:[N]\rightarrow[R]$ chosen as \Theorem{pagh} (again with $z = R$
and $c = 2$). Then, we
let $I = [R]\backslash h_2(L')$. The remaining parts of the algorithm
remain the same.

There are several issues we must address to show that our universe
reduction step still maintains correctness.  Informally, we need that
(a) any $i$ which is a heavy hitter
for $y$ should have exactly one $j\in[n]$ with $h_1(j) = i$ such that
$j$ was a heavy hitter for $x$, 
(b) if
$i$ is a heavy hitter for $x$, then $h_1(i)$ is a heavy hitter for
$y$, and $|y_{h_1(i)}|^p = (1\pm O(\eps))|x_i|^p$ so that $x_i$'s
contribution to $\|x\|_p^p$ is properly approximated by $\highend$,
(c) $\|y\|_p^p = O(\|x\|_p^p)$ with large probability, since the error
term in $\highend$ is $O(\eps\cdot \|y\|_p^p)$,
and (d) the amount of $F_p$ mass not
output by $\lightest$ because it collided with a heavy hitter
for $x$ under $h_1$ is negligible.  Also, the composition $h =
h_1\circ h_2$ for $\lightest$ does not satisfy the conditions of
\Theorem{pagh} even
though $h_1$ and $h_2$ might do so individually.  To see why, as a
simple analogy,
consider that the composition of two purely random functions is no longer
random.  For example, as the number of compositions
increases, the probability of two items colliding increases as
well. Nevertheless, the analysis of $\lightest$ carries over
essentially unchanged in this setting, since whenever considering
the distribution of where two items land under $h$, we can first
condition on them not colliding under $h_1$. Not colliding under $h_1$
happens with $1 -
O(\eps^{18})$ probability, and thus the probability that two items
land in two particular buckets $j,j'\in [R]$ under $h$ is still $(1\pm
o(\eps))/R^2$.

We now give our full description and analysis.  We pick $h_1$ 
as in \Theorem{pagh} with $z=R$ and $c  = c_h$ a sufficiently large
constant. We also
pick $\sigma$ from an $\Omega(\log N)$-wise independent family.
We run an instantiation of $\fphh$ for the vector $y$ with $\phi =
\eps^2/(34C)$ for a sufficiently large constant $C>0$.
We also obtain a value $\tilde{F}_p \in [F_p/2, 3F_p/2]$ using the
algorithm of \cite{KNW10b}. We define $L'$ to be the sublist of those
$w$
output by our $\fphh$ instantiation such that $|\tilde{y}_w|^p \ge
(2\eps^2/7)\tilde{F}_p$.

For ease of presentation in what follows, define $L_\phi$ to
be the list of
$\phi$-heavy hitters of $x$ with respect to $F_p$ (``$L$'', without a
subscript, always
denotes the $\eps^2$-heavy hitters with respect to $x$), and
define $z_i
 = \sum_{w\in h_1^{-1}(i)\backslash L_{\eps^8}}
\sigma(w)x_w$, i.e. $z_i$ is the contribution to $y_i$ from the
significantly light elements of $x$.

\begin{lemma}\LemmaName{khintchine-tail}
For $x\in \R^n$, $\lambda > 0$ with $\lambda^2$ a multiple of $8$, and
random
$z\in\{-1,1\}^n$ drawn from a $(\lambda^2/4)$-wise independent
family, $\Pr[|\inprod{x,z}| > \lambda\|x\|_2] <
2^{-\lambda^2/4}$.
\end{lemma}
\begin{proof}
By Markov's inequality on the random variable
$\inprod{x,z}^{\lambda^2/4}$, $\Pr[|\inprod{x,z}| > \lambda] <
\lambda^{-{\lambda^2/4}}\cdot \E[\inprod{x,z}^{\lambda^2/4}]$.  The claim
follows by applying \Lemma{khintchine}.
\end{proof}

\begin{lemma}\LemmaName{ybounded-lp}
For any $C>0$, there exists $\eps_0$ such that for
$0<\eps < \eps_0$,
$\Pr[\|y\|_p^p > 17C\|x\|_p^p] < 2/C$.
\end{lemma}
\begin{proof}
Condition on $h_1$. 
Define $Y(i)$ to be the vector $x_{h_1^{-1}(i)}$. For any vector $v$
we have $\|v\|_2 \le \|v\|_p$ since $p < 2$.  Letting
$\mathcal{E}$ be the event that no $i\in[N]$
has $|y_i| > 4\sqrt{\log N}\|Y(i)\|_p$, we have
$\Pr[\mathcal{E}] \ge 1 - 1/N^4$
by \Lemma{khintchine-tail}. For $i\in [N]$, again by
\Lemma{khintchine-tail} any
$i\in [N]$ has $|y_i| \le 2t \cdot
\|Y(i)\|_2\le 2t \cdot
\|Y(i)\|_p$ with probability at
least $1 - \max\{1/(2N),2^{-t^2}\}$. Then for fixed $i\in [N]$,
\begin{align*}
\E[|y_i|^p\mid \mathcal{E}] & \le 
2^p\|Y(i)\|_p^p + 
\sum_{t=0}^\infty \Pr\left[(2\cdot 2^t)^p\|Y(i)\|_p^p <
  |y_i|^p \le
(2\cdot 2^{t+1})^p\|Y(i)\|_p^p\mid\mathcal{E}\right] \cdot
(2\cdot
2^{t+1})^p\|Y(i)\|_p^p\\
&\le 2^p\|Y(i)\|_p^p + (1/\Pr[\mathcal{E}])\cdot
\sum_{t=0}^{\log(2\sqrt{\log N})} 2^{-2^{2t}} \cdot
(2\cdot 2^{t+1})^p\|Y(i)\|_p^p\\
&< 4\|Y(i)\|_p^p + (1/\Pr[\mathcal{E}])\cdot
\sum_{t=0}^{\log(2\sqrt{\log N})} 2^{-2^{2t}} \cdot
(2\cdot 2^{t+1})^2\|Y(i)\|_p^p\\
&< 17\|Y(i)\|_p^p
\end{align*}
since
$\Pr[\mathcal{E}] \ge 1 - 1/N^4$ and $\eps_0$ is sufficiently small.
Thus by linearity of
expectation, $\E[\|y\|_p^p\mid\mathcal{E}] \le 17\|x\|_p^p$, which
implies
$\|y\|_p^p\le 17C\|x\|_p^p$ with probability $1 - 1/C$, conditioned on
$\mathcal{E}$
holding. We conclude by again using $\Pr[\mathcal{E}] \ge 1
- 1/N^4$.
\end{proof}

\begin{lemma}\LemmaName{bounded-lp}
With probability at least $1 -
\poly(\eps)$ over $\sigma$,  simultaneously
for all $i\in[N]$ we have that $|z_i| = O(\sqrt{\log(1/\eps)}\cdot
\eps^{6/p}\|x\|_p)$.
\end{lemma}
\begin{proof}
By \Lemma{khintchine-tail}, any individual $i\in [N]$ has $|z_i| \le
4\sqrt{\log(1/\eps)} \cdot
(\sum_{w\in h_1^{-1}(i)\backslash L_{\eps^8}} |x_w|^2)^{1/2}$ with probability at
least $1 - 1/N^4$.  We then apply a union bound and use the
fact that $\ell_p \le \ell_2$ for $p < 2$, so that $|z_i| \le
4\sqrt{\log(1/\eps)} \cdot
(\sum_{w\in h_1^{-1}(i)\backslash L_{\eps^8}} |x_w|^p)^{1/p}$ (call this
event $\mathcal{E}$) with
probability $1 - \poly(\eps)$.

We now prove our lemma, i.e. we show that with
probability $1 - \poly(\eps)$,
$|z_i|^p = O(\log^{p/2}\eps^6\|x\|_p^p)$
simultaneously for all $i\in[N]$. We  apply
\Lemma{good-moment}. Specifically, fix an $i\in [N]$. For all $j$ with
$|x_j|^p \le \eps^8\|x\|_p^p$, let $X_j
= |x_j|^p\cdot \indicate_{h_1(j) = i}$. Then, in
the notation of \Lemma{good-moment},
$\mu_j = |x_j|^p/N$, and $\sigma_j^2 \le |x_j|^{2p}/N$, and thus $\mu
= \|x\|_p^p/N$ and $\sigma^2 \le \|x\|_{2p}^{2p}/N \le
\eps^8\|x\|_p^p/N$. Also, $K = \eps^8\|x\|_p^p$.
Then if $h_1$ were $\ell$-wise independent for $\ell = 10$,
\Lemma{good-moment} would give
$$ \Pr\left[\left|\sum_i X_i - \|x\|_p^p/N\right| >
  \eps^6\|x\|_p^p\right] < 2^{O(\ell)}\cdot (\eps^{7\ell} +
\eps^{2\ell}) = O(\eps/N) .$$
A union bound would then give that with probability $1-\eps$, the $F_p$
mass in any bucket from items $i$ with
$|x_i|^p\le \eps^8\|x\|_p^p$ is at most
$\eps^6\|x\|_p^p$. 
Thus by a union bound with event $\mathcal{E}$,
$|z_i|^p = O(\log^{p/2}\eps^6\|x\|_p^p)$ for all $i\in[N]$
with probability $1 - \poly(\eps)$.

Though, $h_1$ is not $10$-wise independent.  Instead, it is selected
as in \Theorem{pagh}.  However, for any constant
$\ell$, by increasing the constant $c_h$ in our definition of $h_1$ we
can ensure that our $\ell$th moment bound for $(\sum_i X_i -
\mu)$ is
preserved to within a constant factor, which is sufficient to apply
\Lemma{good-moment}.
\end{proof}

\begin{lemma}\LemmaName{still-heavy}
With probability $1 -
\poly(\eps)$, for all
$w\in L$ we have
$|y_{h_1(w)}|^p = (1\pm O(\eps))|x_w|^p$, and thus with probability $1
- \poly(\eps)$ when conditioned on $\|y\|_p^p \le 17C\|x\|_p^p$, we
have that if $w$ is
an $\alpha$-heavy hitter for $x$, then
$h_1(w)$ is an $\alpha/(34C)$-heavy hitter for $y$.
\end{lemma}
\begin{proof}
Let $w$ be in $L$.
We know from \Lemma{bounded-lp} that $|z_{h_1(w)}| \le
2\sqrt{\log(1/\eps)}\eps^{6/p}\|x\|_p$ with probability $1 -
\poly(\eps)$, and that the elements of $L$ are perfectly hashed
under $h_1$ with probability $1- \poly(\eps)$.  Conditioned on this
perfect hashing, we have that $|y_{h_1(w)}| \ge |x_w| -
2\eps^{6/p}\sqrt{\log(1/\eps)}\|x\|_p$. Since for $w\in L$ we have
$|x_w| \ge
\eps^{2/p}\|x\|_p$, and since $p< 2$, we have $|y_{h_1(w)}| \ge (1 -
O(\eps))|x_w|$.

For the second part of the lemma,
$(1 - O(\eps))|x_w| > |x_w|/2^{1/p}$ for $\eps_0$ sufficiently
small. Thus if $w$ is an $\alpha$-heavy hitter for $x$, then
$h_1(w)$ is an $\alpha/(34C)$-heavy hitter for $y$.
\end{proof}

Finally, the following lemma follows from a Markov bound followed by a
union bound.

\begin{lemma}\LemmaName{small-noise}
For $w\in [n]$ consider the quantity $s_w = \sum_{\substack{v\neq
    w\\h(v) = h(w)}} |x_v|^p$. Then, with
probability at least $1-O(\eps)$, $s_w \le \eps^{15}\|x\|_p^p$
simultaneously for all $w\in L$.
\end{lemma}

We now put everything together.
We set $\epsilon =
\eps\log(1/\eps)$. As stated earlier, we define $L'$ to be the sublist
of those $w$
output by our $\fphh$ instantiation with $\phi = \epsilon^2$
such that $|\tilde{y}_w|^p \ge
(2\eps^2/7)\tilde{F}_p$.  We
interpret updates to $x$ as updates to $y$ to then be fed into
$\highend$, with $\alpha = \epsilon^2/(34C)$. 
Thus both $\highend$ and
$\fphh$ require $O(\eps^{-2}\log(nmM/\eps))$ space. 
We now define some events.

\vspace{.1in}

\noindent {\bf Event $\mathcal{A}$}. $L_{\eps^8}$
is perfectly hashed
under $h_1$, and $\forall i\in [N], |z_i|^p = O(\log(1/\eps)^{p/2}\cdot
\eps^6\|x\|_p^p)$.

\vspace{.1in}

\noindent {\bf Event $\mathcal{B}$}. $\forall w\in L_{\epsilon^2}$,
$h_1(w)$ is
output as an $\epsilon^2/(34C)$-heavy hitter by $\fphh$.

\vspace{.1in}

\noindent {\bf Event $\mathcal{C}$}. $\forall w\in
L_{\epsilon^2/18}$,
$|y_{h_1(w)}|
= (1\pm O(\eps))|x_w|$.

\vspace{.1in}

\noindent {\bf Event $\mathcal{D}$}. $\tilde{F}_p \in [(1/2)\cdot
\|x\|_p^p, (3/2)\cdot\|x\|_p^p]$, and $\highend$, $\lightest$, and
$\fphh$ succeed.

\vspace{.1in}

Now, suppose $\mathcal{A}$, $\mathcal{B}$, $\mathcal{C}$, and
$\mathcal{D}$ all occur. Then for $w\in L_{\epsilon^2}$, 
$w$ is output by $\fphh$, and furthermore $|y_{h_1(w)}|^p \ge
(1-O(\eps))|x_w|^p \ge |x_w|^p/2 \ge \epsilon^2\|x\|_p^p/2$. Also,
$\tilde{y}_{h_1(w)}^p \ge
(6/7)\cdot |y_{h_1(w)}|^p$. Since $\tilde{F}_p \le 3\|x\|_p^p/2$, we
have that $h_1(w)\in L'$. Furthermore, we also know that for $i$
output
by $\fphh$, $\tilde{y}_i^p \le (9/7)\cdot |y_i|^p$, and thus $i\in
L'$ implies $|y_i|^p \ge (\epsilon^2/9)\cdot \|x\|_p^p$. Notice that
by event $\mathcal{A}$, each $y_i$ is $z_i$, plus potentially $x_{w(i)}$
for some $x_{w(i)}\in L_{\eps^8}$. If $|y_i|^p \ge (\epsilon^2/9)\cdot
\|x\|_p^p$, then there must exist such a $w(i)$, and furthermore it
must be that $|x_{w(i)}|^p \ge (\epsilon^2/18)\cdot \|x\|_p^p$.  Thus,
overall, $L'$ contains $h_1(w)$ for all $w\in L_{\epsilon^2}$, and
furthermore if $i\in L'$ then $w(i)\in L_{\epsilon^2/18}$.

Since $L'$ contains $h_1(L_{\epsilon^2})$,
$\lightest$ outputs $\|x_{[n]\backslash h^{-1}(L')}\|_p^p \pm
O(\eps\|x\|_p^p)$. Also, $\highend$ outputs $\|y_{L'}\| \pm
O(\eps)\cdot 
\|y\|_p^p$. Now we analyze correctness.  We have
$\Pr[\mathcal{A}] = 1 - \poly(\eps)$, $\Pr[\mathcal{B}\ |\ \|y\|_p^p
\le 17C\|x\|_p^p] = 1 - \poly(\eps)$, $\Pr[\mathcal{C}] = 1 -
\poly(\eps)$, and $\Pr[\mathcal{D}] \ge 5/8$.  We also have
$\Pr[\|y\|_p^p \le 17C\|x\|_p^p] \ge 1 - 2/C$.  Thus by a union bound
and setting $C$ sufficiently large, we have $\Pr[\mathcal{A}\wedge
\mathcal{B}\wedge \mathcal{C}\wedge \mathcal{D}\wedge (\|y\|_p^p \le
17C\|x\|_p^p)] \ge 9/16$. 
Define $L_{\mathrm{inv}}$ to
be the set $\{w(i)\}_{i\in L'}$, i.e. the
heavy hitters of $x$ corresponding to the heavy hitters in $L'$ for
$y$.
Now, if all these events occur, then
$\|x_{[n]\backslash h^{-1}(L')}\|_p^p = \|x_{[n]\backslash
  L_{\mathrm{inv}}}\|_p^p \pm O(\eps^{15})\|x\|_p^p$ with probability
  $1 - O(\eps)$ by \Lemma{small-noise}. We also have, since
  $\mathcal{C}$ occurs and conditioned on $\|y\|_p^p = O(\|x\|_p^p)$,
  that $\|y_{L'}\| \pm O(\eps)\cdot \|y\|_p^p =
  \|x_{L_{\mathrm{inv}}}\|_p^p \pm O(\eps)\cdot \|x\|_p^p$. Thus,
  overall, our algorithm outputs $\|x\|_p^p \pm O(\eps)\cdot
  \|x\|_p^p$  with probability $17/32 >
  1/2$ as desired. Notice this probability can be amplified to
  $1-\delta$ by
  outputting the median of $O(\log(1/\delta))$ independent
  instantiations.

We further note that for a single instantiation of $\lightest$, we
have $\E_h[\Var_s[\Phi']] = O(\epsilon^2\|x\|_p^{2p})$. Once $h$ is
fixed, the variance of $\Phi'$ is simply the sum of variances across
the $D_j$ for $j\notin h_1(L')$. Thus, it suffices for the $D_j$ to
use pairwise independent randomness.  Furthermore, in repeating
$O((\epsilon/\eps)^2)$ parallel repetitions of $\lightest$, it
suffices that all the $D_j$ across all parallel repetitions use
pairwise independent randomness, and the hash function $h$ can remain
the same. Thus, as discussed in
\Remark{modified}, the coefficients of the degree-$O(1/\eps^p)$
polynomials used in all $D_j$ combined can be generated by just two
coefficient vectors, and thus the update time of $\lightest$ with
$O((\epsilon/\eps)^2)$ parallel repetitions is just
$O((\epsilon/\eps)^2 + O(\log^2(1/\eps)\log\log(1/\eps))) =
O(\log^2(1/\eps)\log\log(1/\eps))$.
Thus overall, we have the
following theorem.

\begin{theorem}
There exists an algorithm such that given $0<p<2$ and $0<\eps<1/2$,
the algorithm outputs $(1\pm \eps)\|x\|_p^p$ with probability $2/3$
using $O(\eps^{-2}\log(nmM/\eps))$ space. The update time is
$O(\log^2(1/\eps)\log\log(1/\eps))$. The reporting time is
$O(\eps^{-2}\log^2(1/\eps)\log\log(1/\eps))$.
\end{theorem}

The space bound above can be assumed $O(\eps^{-2}\log(mM) +
\log\log n)$ by comments in \Section{notation}.


\bibliographystyle{plain}

\bibliography{allpapers}

\newpage
\appendix

\section{Appendix}
\subsection{A heavy hitter algorithm for $F_p$}\SectionName{fp-hh}
Note that $\fpreport$, $\fpupdate$,
and $\fpspace$ below can be as in the statement in \Section{heavy-contrib}
by using the algorithm of \cite{KNW10b}.

\vspace{.1in}

\noindent \BoldTheorem{fp-hh}
{\it
There is an algorithm $\fphh$ satisfying the following properties.
Given $0<\phi,\delta<1/2$ and black-box access to an
$F_p$-estimation algorithm $\fpest(\eps',\delta')$ with $\eps' = 1/7$
and $\delta' =
\phi\delta/(12(\log(\phi n) + 1))$, 
$\fphh$ produces a list $L$ such that $L$
contains all $\phi$-heavy hitters and does not contain indices which
are not $\phi/2$-heavy hitters with probability
at least $1-\delta$. For each $i\in L$, the algorithm also
outputs $\sign(x_i)$, as well as an estimate $\tilde{x}_i$ of $x_i$
satisfying $\tilde{x}_i^p \in [(6/7)|x_i|^p, (9/7)|x_i|^p]$.
Its space usage is $O(\phi^{-1}\log (\phi n)\cdot
\fpspace(\eps',\delta') + \phi^{-1}\log(1/(\delta
\phi))\log(nmM))$. Its update
time is $O(\log
(\phi n)\cdot \fpupdate(\eps', \delta') + \log(1/(\delta\phi)))$. Its
reporting time is 
$O(\phi^{-1}(\log (\phi n)\cdot \fpreport(\eps', \delta') +
\log(1/(\delta\phi))))$.
Here,  $\fpreport(\eps',\delta')$, $\fpupdate(\eps',\delta')$, and
$\fpspace(\eps',\delta')$ are the reporting time, update time, and space
consumption of $\fpest$ when a $(1\pm\eps')$-approximation to $F_p$ is
desired with probability at least $1-\delta'$.
}

\begin{proof}
First we argue with $\delta' = \phi\delta/(12(\log n + 1))$.
We assume without loss of generality that $n$ is a power of $2$.
Consider the following data structure $\basicfphh(\phi', \delta,
\eps', k)$, where $k\in\{0,\ldots,\log n\}$.  We set $R =
\ceil{1/\phi'}$ and pick a 
function $h:\{0,\ldots,2^k-1\}\rightarrow[R]$ at random from a
pairwise independent
hash family. We also create instantiations $D_1,\ldots,D_R$ of
$\fpest(\eps',1/5)$. This entire structure is then repeated
independently in parallel $T =
\Theta(\log(1/\delta))$ times, so that we have hash functions
$h_1,\ldots,h_T$, and instantiations $D_i^j$ of $\fpest$ for $i,j\in
[R]\times [T]$.  For an integer $x$ in $[n]$, let $\prefix(x, k)$
denote
the length-$k$ prefix of $x-1$ when written in binary, treated as an
integer in $\{0,\ldots,2^k-1\}$.
Upon receiving an update $(i,v)$ in the stream, we feed this update to
$D_{h_j(\prefix(i,k))}^j$ for each $j\in[T]$. 

For $t\in \{0,\ldots,2^k-1\}$,
let  $F_p(t)$ denote the $F_p$ value of the vector $x$ restricted to
indices $i\in[n]$ with $\prefix(i) = t$.
Consider the procedure $\query(t)$
which outputs the median of $F_p$-estimates given
by $D_{h_j(t)}^j$ over all $j\in[T]$. 
We now argue that the output of
$\query(t)$ is in the interval $[(1-\eps')\cdot F_p(t),
(1+\eps')\cdot(F_p(t)
+ 5\phi'\|x\|_p^p)]]$, i.e. $\query(t)$ ``succeeds'', with probability
at least $1-\delta$.

For any $j\in[T]$, consider the
actual $F_p$ value $F_p(t)^j$ of the vector $x$ restricted to
coordinates $i$ such that $h_j(\prefix(i,k)) = h_j(t)$. Then $F_p(t)^j
= F_p(t) + R(t)^j$, where $R(t)^j$ is the
$F_p$ contribution of the $i$ with $\prefix(i,k)\neq t$, yet
$h_j(\prefix(i,k)) = h(t)$.
We have
$R(t)^j \ge 0$ always, and furthermore
$\E[R(t)^j] \le \|x\|_p^p/R$ by pairwise independence of $h_j$. Thus
by Markov's inequality, $\Pr[R(t)^j > 5\phi'\|x\|_p^p] <
1/5$. Note for any fixed $j\in[T]$, the $F_p$-estimate output by
$D_{h(t)}^j$ is in $[(1-\eps')\cdot F_p(t),
(1+\eps')\cdot(F_p(t)
+ 5\phi'\|x\|_p^p)]]$ as long as both the events ``$D_{h(t)}^j$
successfully gives a $(1\pm\eps')$-approximation'' and ``$R(t)^j \le
5\phi'\|x\|_p^p$'' occur. This happens with probability at least
$3/5$. Thus, by a Chernoff bound, the output of $\query(t)$ is in
the desired interval
with probability at least $1-\delta$.

We now define the final $\fphh$ data structure.
We maintain one global instantiation $D$ of $\fpest(1/7,\delta/2)$.
We also use the dyadic
interval idea for $L_1$-heavy hitters given in
\cite{CM05}. Specifically, we imagine building a binary tree
$\mathcal{T}$ over the
universe $[n]$ (without loss of generality assume $n$ is a power of
$2$). The number of levels in the tree is $\ell = 1 + \log n$, where
the root
is at level $0$ and the leaves are at level $\log n$. For each level
$j\in\{0,\ldots, \ell\}$, we maintain an instantiation
$B_j$ of $\basicfphh(\phi/80, \delta', 1/7, j)$ for $\delta'$ as in
the theorem statement. When we receive an
update $(i,v)$ in the stream, we feed the update to $D$ and also to
each $B_j$.

We now describe how to answer a query to output the desired list
$L$. 
We first query $D$ to obtain $\tilde{F}_p$, an approximation to
$F_p$. We next initiate an iterative procedure on our binary tree,
beginning at the root, which proceeds level by level.  The procedure
is as follows.  Initially, we
set $L = \{0\}$, $L' = \emptyset$, and $j = 0$.  For each $i\in L$, we
perform $\query(i)$ on $B_j$ then add $2i$ and $2i+1$ to $L'$ if the
output of $\query(i)$ is at least $3\phi\tilde{F}_p/4$. After
processing every $i\in L$, we then set $L\leftarrow L'$ then
$L'\leftarrow \emptyset$, and we increment $j$. This continues until
$j=1+\log n$, at which point we halt and return $L$. 
We now show why
the list $L$ output by this procedure satisfies the claim in the
theorem statement. We condition on the event $\mathcal{E}$
that $\tilde{F}_p = (1\pm 1/7)F_p$, and also on the event
$\mathcal{E}'$ that every query made throughout the recursive
procedure is successful. Let $i$ be such that $|x_i|^p \ge \phi F_p$.
Then, since $F_p(\prefix(i, j)) \ge |x_i|^p$
for any $j$, we always have that $\prefix(i,j)\in L$ at the end of the
$j$th round of our iterative procedure, since $(6/7)|x_i|^p \ge
(3/4)\phi\tilde{F}_p$ given $\mathcal{E}$.  Now, consider an $i$ such
that $|x_i|^p < (\phi/2)F_p$. Then, $(8/7)\cdot (|x_i|^p - 5\cdot
(\phi/80)) < 3\phi\tilde{F}_p/4$, implying $i$ is not included in the
final output list. Also, note that since the query at the leaf
corresponding to $i\in L$ is successful, then by definition of a
successful query, we are given an estimate $\tilde{x}_i^p$ of
$|x_i|^p$ by the corresponding $\basicfphh$ structure satisfying
$\tilde{x}_i^p \in [(6/7)|x_i|^p, (8/7)|x_i|^p + (\phi/16)F_p]$, which
is $[(6/7)|x_i|^p, (9/7)|x_i|^p]$ since $|x_i|^p \ge (\phi/2)F_p$.


We now only need to argue that $\mathcal{E}$ and $\mathcal{E}'$ occur
simultaneously with large probability. We have $\Pr[\mathcal{E}] \ge 1 -
\delta/2$. For $\mathcal{E}'$, note there are at most $2\phi$
$\phi/2$-heavy hitters at any level of the tree, where at level $j$
we are referring to heavy hitters of the $2^j$-dimensional vector
$y_j$ satisfying $(y_j)_i^p = \sum{\prefix(t, j) = i} |x_t|^p$.  As
long as the
$\query(\cdot)$ calls made for all $\phi/2$-heavy hitters and their
two children throughout the tree succeed (including at the root),
$\mathcal{E}'$ holds.  Thus, $\Pr[\mathcal{E}'] \ge 1 - \delta'\cdot
6(\log n + 1)\phi^{-1} = 1 - \delta/2$. Therefore, by a union bound
$\Pr[\mathcal{E} \wedge \mathcal{E}'] \ge 1 - \delta$.

Finally, notice that the number of levels in $\fphh$ can be reduced
from $\log n$ to $\log n - \log \ceil{1/\phi} = O(\log (\phi n))$ by
simply ignoring the top $\log \ceil{1/\phi}$ levels of the tree. Then,
in the topmost level of the tree which we maintain, the universe size
is $O(1/\phi)$, so we can begin our reporting procedure by querying
all these universe items to determine which subtrees to recurse upon.

To recover $\sign(x_w)$ for each $w\in L$, we use the $\countsketch$
data structure of \cite{CCF02} with $T = (21\cdot 2^p)/\phi$ columns and
$C = \Theta(\log(1/(\delta\phi)))$ rows; the space is
$O(\phi^{-1}\log(1/(\delta\phi))\log(nmM))$, and the update time is
$O(\log(1/(\delta\phi)))$. $\countsketch$ operates by, for each row $i$,
having a pairwise independent hash function $h_i:[n]\rightarrow [T]$ and
a $4$-wise independent hash function $\sigma_i:[n]\rightarrow
\{-1,1\}$. There are $C\cdot T$ counters $A_{i,j}$ for
$(i,j)\in[C]\times [T]$. Counter $A_{i,j}$ maintains $\sum_{h_i(v) =
  j} \sigma_i(v)\cdot x_v$. For $(i,j)\in [C]\times [T]$, let
$x^{i}$ be the vector $x$ restricted to coordinates $v$ with $h_i(v)
= h_i(w)$, other than $w$ itself.  Then for fixed $i$, the expected
contribution to
$\|x^i\|_p^p$ is at most
$\|x\|_p^p/T$, and thus is at most $10\|x\|_p^p/T$ with
probability
$9/10$ by Markov's inequality. Conditioned on this event, $|x_w| >
\|x^i\|_p/2 \ge \|x^i\|_2/2$. The analysis of $\countsketch$ also
guarantees $|A_{i,h_i(w)} - \sigma_i(w)x_w| \le 2\|x^i\|_2$ with
probability at least $2/3$, and thus by a union bound, $|x_w| >
|A_{i,h_i(w)} - \sigma_i(w)x_w|$ with probability at least $11/20$, in
which case $\sigma_i(w)\cdot \sign(A_{i,h_i(w)}) = \sign(x_w)$.  Thus,
by a Chernoff
bound over all rows, together with a union bound over all $w\in L$, we
can recover $\sign(x_w)$ for all $w\in L$ with probability $1 -
\delta$.
\end{proof}

\subsection{Proof of \Theorem{gme-good}}\SectionName{gme}
In this section we prove \Theorem{gme-good}. The data structure and
estimator we give is a 
slightly modified version of the geometric mean estimator of Li
\cite{Li08b}. Our modification allows us to show that only bounded
independence is required amongst the
$p$-stable random variables in our data structure.
Before giving our $D_p$ and $\Est_p$, we first define the
{\em $p$-stable distribution}.

\begin{definition}[Zolotarev
  \cite{Zolotarev86}]\DefinitionName{pstable}
For $0<p<2$, there exists a probability distribution $\mathcal{D}_p$
called the {\em $p$-stable distribution} satisfying the following
property. For any positive integer $n$ and vector $\x\in\mathbb{R}^n$,
if
$Z_1,\ldots,Z_n \sim \mathcal{D}_p$ are independent, then
$\sum_{j=1}^n Z_j \x_j \sim \|\x\|_pZ$ for $Z\sim\mathcal{D}_p$.
\end{definition}

Li's {\em geometric mean estimator} is as follows. 
For some positive integer $t > 2$, select a matrix $A\in\R^{t\times
  n}$ with independent $p$-stable entries, and maintain $y = Ax$ in
the stream. Given $y$, the estimate of $\|x\|_p^p$ is then
$C_{t,p}\cdot (\prod_{j=1}^t |y_j|^{p/t})$ for some constant
$C_{t,p}$. For \Theorem{gme-good}, we make the
following adjustments.  First, we require $t > 4$. Next, 
for any fixed row of $A$ we only require that the entries be
$\Omega(1/\eps^p)$-wise independent, though the rows
themselves we keep independent. Furthermore, in parallel we run the
algorithm of \cite{KNW10b} with constant error parameter to obtain a value
$\tilde{F}_p$ in $[\|x\|_p^p/2, 3\|x\|_p^p/2]$.
The $D_p$ data structure of
\Theorem{gme-good} is then simply $y$, together with the state
maintained by the algorithm of \cite{KNW10b}.
The estimator $\Est_p$ is $\min\{C_{t,p}\cdot (\prod_{j=1}^t
|y_j|^{p/t}), \tilde{F}_p/\eps\}$.  To state the value $C_{t,p}$, we
use the following theorem.

\begin{theorem}[{\cite[Theorem 2.6.3]{Zolotarev86}}]\TheoremName{zolotarev}
For $Q\sim\mathcal{D}_p$ and $-1<\lambda < p$,
$$\E[|Q|^\lambda] =
\frac 2\pi\Gamma\left(1 -
\frac
{\lambda}{p}\right)\Gamma(\lambda)\sin\left(\frac{\pi}{2}\lambda\right)
.$$
\end{theorem}

\Theorem{zolotarev} implies that we should set 
$$C_{t,p} = \left[\frac 2{\pi}\cdot \Gamma\left(1 - \frac
    1t\right)\cdot \Gamma\left(\frac pt\right)\cdot
  \sin\left(\frac{\pi p}{2t}\right)\right]^{-t} .$$

To carry out our analysis, we will need the following theorem,
which gives a way of producing a smooth approximation of
the indicator function of an interval while maintaining good bounds on
high order derivatives.

\begin{theorem}[{\cite{DKN10}}]\LemmaName{ftmol}
For any interval $[a,b]\subseteq \R$ and integer $c>0$, there exists
a nonnegative function $\tilde{I}^c_{[a,b]}:\R \rightarrow\R$
satisfying the
following properties:
\begin{enumerate}
\item[i.] $\|(\tilde{I}^c_{[a,b]})^{(\ell)}\|_\infty \le (2c)^\ell$ for all
  $\ell\ge 0$.
\item[ii.] For any $x\in\R$, $|\tilde{I}^c_{[a,b]}(x) - I_{[a,b]}(x)| \le
  \min\{1, 5/(2c^2\cdot d(x,\{a,b\})^2)\}$.
\end{enumerate}
\end{theorem}

We also need the following lemma of \cite{KNW10b}, which argues that
smooth, bounded functions have their expectations approximately
preserved when
their input is a linear form evaluated at boundedly independent
$p$-stable random variables, as opposed to completely independent
$p$-stable random variables.

\begin{lemma}[{\cite[Lemma 2.2]{KNW10b}}]\LemmaName{hairy}
There exists an $\eps_0>0$ such that the following holds. Let
$0<\eps<\eps_0$ and $0<p<2$ be given. Let $f:\R\rightarrow\R$ satisfy
$\|f^{(\ell)}\|_\infty = O(\alpha^\ell)$ for all $\ell\ge 0$, for some
$\alpha$ satisfying $\alpha^p \ge \log(1/\eps)$. Let $k =
\alpha^p$. Let $x\in\R^n$ satisfy $\|x\|_p = O(1)$. Let
$R_1,\ldots,R_n$ be drawn from a $3Ck$-wise independent family of
$p$-stable random variables for $C>0$ a sufficiently large constant,
and let $Q$ be the product of $\|x\|_p$
and a $p$-stable random variable.  Then $|\E[f(R)] - \E[f(Q)]| =
O(\eps)$.
\end{lemma}

We now prove a tail bound for linear forms over $k$-wise independent
$p$-stable random variables.  Note that for a random variable
$X$ whose moments are bounded, one has $\Pr[X-\E[X] > t]
\le \E[(X-\E[X])^k] / t^k$ by applying Markov's inequality to the
random variable $(X-\E[X])^k$ for some even integer $k\ge 2$.
Unfortunately, for $0<p<2$, it is known that even the second moment of
$\mathcal{D}_p$ is already infinite, so this method cannot be applied.
We instead prove our tail bound via FT-mollification of 
$I_{[t,\infty)}$, since $\Pr[X \ge t] = \E[I_{[t,\infty)}(X)]$.

We will need to refer to the following lemma.

\begin{lemma}[Nolan {\cite[Theorem 1.12]{Nolan09}}]\LemmaName{nolan}
For fixed $0<p<2$, the probability density function $\varphi_p$ of the
$p$-stable
distribution satisfies $\varphi_p(x) = O(1/(1 + |x|^{p+1}))$ and is an even
function.  The cumulative
distribution function
satisfies $\Phi_p(x) = O(|x|^{-p})$.
\end{lemma}

We now prove our tail bound.

\begin{lemma}\LemmaName{kwise-tailbound}
Suppose $x\in\R^n$, $\|x\|_p = 1$, $0<\eps<1$ is
given, and
$R_1,\ldots,R_n$ are $k$-wise
independent $p$-stable random variables for $k\ge 2$.
Let $Q\sim\mathcal{D}_p$.
Then  for all
$t\ge 0$, $R =
\sum_{i=1}^n R_ix_i$ satisfies
$$ |\Pr[|Q|\ge t] - \Pr[|R| \ge t]| =  O(k^{-1/p}/(1+t^{p+1})
+ k^{-2/p}/(1+t^2) + 2^{-\Omega(k)}) .$$
\end{lemma}
\begin{proof}
We have $\Pr[|Z| \ge t] = \E[I_{[t,\infty)}(Z)] + \E[I_{(-\infty,
  t]}(Z)]$ for any random variable $Z$, and thus we will argue
$|\E[I_{[t,\infty)}(Q)] - \E[I_{[t,\infty)}(R)]| =  O(k^{-1/p}/(1+t^{p+1})
+ k^{-2/p}/(1+t^2) + 2^{-\Omega(k)})$; a similar argument shows the same bound
for $|\E[I_{(-\infty,t]}(Q)] - \E[I_{(-\infty,t]}(R)]|$.

We argue the following chain of inequalities for $c=k^{1/p}/(3C)$, for
$C$ the constant in \Lemma{hairy}, and we define
$\gamma = k^{-1/p}/(1+t^{p+1}) + k^{-2/p}/(1+t^2)$:
$$ \E[I_{[t,\infty)}(Q)] \approx_\gamma
\E[\tilde{I}^c_{[t,\infty)}(Q)]
\approx_{2^{-c^p}} \E[\tilde{I}^c_{[t,\infty)}(R)]
\approx_{\gamma + 2^{-c^p}}
\E[I_{[t,\infty)}(R)] .$$

\noindent $\mathbf{\E[I_{[t,\infty)}(Q)] \approx_\gamma
\E[\tilde{I}^c_{[t,\infty)}(Q)]}$: 
Assume $t \ge 1$.  We have
\begin{align}
\nonumber |\E[I_{[t,\infty)}(Q)] - \E[\tilde{I}^c_{[t,\infty)}(Q)]| & \le
 \E[|I_{[t,\infty)}(Q) - \tilde{I}^c_{[t,\infty)}(Q)|]\\
{} & \le \Pr[|Q-t| \le 1/c] + \left(\sum_{s=1}^{\log(ct) - 1} \Pr[|Q-t| \le
2^s/c]\cdot O(2^{-2s})\right)\EquationName{replace-this}\\
\nonumber{} & \hspace{.2in} + \Pr[|Q-t| > t/2]\cdot O(c^{-2}t^{-2}) \\
\nonumber {} & = O(1/(c\cdot t^{p+1})) + O(c^{-2}t^{-2})
\end{align}
since $\Pr[|Q - t| \le 2^s/c$ is $O(2^s/(c\cdot t^{p+1})$
as long as $2^s/c \le t/2$.

In the case $0 < t < 1$, we repeat the same argument as above but
replace  \Equation{replace-this} with a summation from $s=1$ to
$\infty$, and also remove the additive $\Pr[|Q-t|>t/2]\cdot
O(c^{-2}t^{-2})$ term. Doing so gives an overall upper bound of
$O(1/c)$ in this case.

\vspace{.1in}

\noindent $\mathbf{\E[\tilde{I}^c_{[t,\infty)}(Q)] \approx_{2^{-c^p}}
\E[\tilde{I}^c_{[t,\infty)}(R)]}$: This follows from \Lemma{hairy}
with $\eps = 2^{-c^p}$ and $\alpha = c$.

\vspace{.1in}

\noindent $\mathbf{\E[\tilde{I}^c_{[t,\infty)}(R)] \approx_{\gamma + 2^{-c^p}}
\E[I_{[t,\infty)}(R)]}$: 
We would like to apply the same argument as when showing
$\E[\tilde{I}^c_{[t,\infty)}(Q)] \approx_\gamma
\E[I_{[t,\infty)}(Q)]$ above. The trouble is, we must bound
$\Pr[|R-t| > t/2]$ and $\Pr[|R-t| \le 2^s/c]$ given that the $R_i$ are
only $k$-wise independent. For the first probability, we above only
used that
$\Pr[|Q-t|  > t/2] \le 1$, which still holds with $Q$ replaced by
$R$.  

For the second probability, observe $\Pr[|R-t| \le 2^s/c] =
\E[I_{[t - 2^s/c, t + 2^s/c]}(R)]$. Define $\delta = 2^s/c + b/c$ for
a sufficiently large constant $b>0$ to be determined later. Then,
arguing as above, we have $\E[\tilde{I}^c_{[t-\delta, t+\delta]}(R)]
\approx_{2^{-c^p}} \E[\tilde{I}^c_{[t-\delta,t+\delta]}(Q)]
\approx_\gamma \E[I_{[t-\delta,t+\delta]}(Q)]$, and we also know
$\E[I_{[t-\delta,t+\delta]}(Q)] = O(\E[I_{[t - 2^s/c, t +
  2^s/c]}(Q)]) = O(\Pr[|Q - t| \le 2^s/c]) = O(2^s/(c\cdot
t^{p+1}))$. Now, 
for $x\notin [t-2^s/c, t+2^s/c]$, $I_{[t-2^s/c,t+2^s/c]}(x) = 0$ while 
$I_{[t-\delta,t+\delta]}(x) =1$. For $x\in [t-2^s/c,t+2^s/c]$, the
distance from $x$ to $\{t-\delta,t+\delta\}$ is at least $b/c$,
implying $\tilde{I}^c_{[t-\delta,t+\delta]}(x) \ge 1/2$ for $b$
sufficiently large by item (ii) of \Lemma{ftmol}. Thus, $2\cdot
\tilde{I}^c_{[t-\delta,t+\delta]} \ge I_{[t-2^s/c,t+2^s/c]}$ on $\R$,
and thus in particular, $\E[I_{[t-2^s/c,t+2^s/c]}(R)] \le 2\cdot
\E[\tilde{I}^c_{[t-\delta,t+\delta]}(R)]$. Thus, in summary,
$\E[I_{[t-2^s/c,t+2^s/c]}(R)] = O(2^s/(c\cdot t^{p+1}) + \gamma +
2^{-c^p})$.
\end{proof}



We now prove the main lemma of this section, which implies
\Theorem{gme-good}.

\begin{lemma}
Let $x\in\R^n$ be such that $\|x\|_p = 1$, and suppose
$0<\eps<1/2$. Let $0<p<2$, and let $t$ be any constant
greater than $4/p$.
Let $R_1,\ldots,R_n$ be
$k$-wise independent $p$-stable random variables for $k =
\Omega(1/\eps^p)$, and let $Q$ be a
$p$-stable random variable.  Define $f(x) = \min\{|x|^{1/t}, T\}$,
for
$T = 1/\eps$.
Then, $|\E[f(R)] - \E[|Q|^{1/t}] = O(\eps)$ and $\E[f^2(R)] =
O(\E[|Q|^{2/t}])$.
\end{lemma}
\begin{proof}
We first argue $|\E[f(R)] - \E[|Q|^{1/t}] = O(\eps)$.
We argue through the chain of inequalities
$$ \E[|Q|^{1/t}] \approx_{\eps} \E[f(Q)] \approx_\eps
\E[f(R)] .$$

\vspace{.1in}

\noindent $\mathbf{\E[|Q|^{1/t}] \approx_{\eps}
\E[f(Q)]}$: We have
\begin{align*}
 |\E[|Q|^{1/t}] - \E[f(Q)]| & = 2\int_{T^t}^\infty (x^{1/t} - T)\cdot
\varphi_p(x)dx \\
{} & = \int_{T^t}^\infty (x^{1/t} - T)\cdot O(1/x^{p+1})dx\\
{} & = O\left(T^{1 - tp}\cdot \left(\frac{t}{pt-1}+
    \frac{1}{p}\right)\right)\\
{} & = O(1/(Tp))\\
{} & = O(\eps)
\end{align*}

\noindent $\mathbf{\E[f(Q)] \approx_{\eps}
\E[f(R)]}$:
Let $\varphi_p^+$ be the probability density function corresponding to the
distribution of $|Q|$, and let $\Phi_p^+$ be its cumulative distribution
function. 
Then, by integration by parts and \Lemma{kwise-tailbound},
\begin{align}
\nonumber \E[f(Q)] & = \int_{0}^{T^t} x^{1/t}\varphi_p^+(x)dx + T\cdot\int_{T^t}^\infty
\varphi_p^+(x)dx\\
\nonumber {} & = -[x^{1/t} \cdot (1 - \Phi_p^+(x))]_0^{T^t} - T\cdot [(1 -
\Phi_p^+(x))]_{T^t}^\infty + \frac 1t\int_0^{T^t}\frac{1}{x^{1-1/t}}(1 -
\Phi_p^+(x))dx \\
\nonumber {} & = \frac 1t\int_0^{T^t}\frac{1}{x^{1-1/t}}\cdot\Pr[|Q|\ge x] dx\\
\nonumber {} & = \frac 1t\int_0^{T^t}\frac{1}{x^{1-1/t}}\cdot (\Pr[|R|\ge x] +
O(k^{-1/p}1/(1+x^{p+1}) + k^{-2/p}/(1+x^2) + 2^{-\Omega(k)}))dx\\
\nonumber {} & = \E[f(R)] + \int_0^1x^{1/t - 1}\cdot O(k^{-1/p} +
k^{-2/p} + 2^{-\Omega(k)}))dx\\
{} &\hspace{.2in} +
\int_1^{T^t}x^{1/t - 1}\cdot O(k^{-1/p}/x^{p+1} + k^{-2/p}/x^2
+ 2^{-\Omega(k)}))dx \EquationName{geometric-error}\\
\nonumber {} & = \E[f(R)] + O(\eps) + O\left(\frac{1}{k^{1/p}}\cdot
\left(\frac{1}{T^{tp + t-1}} - 1\right)\cdot \frac{1}{\frac 1t - p
  -1}\right)\\
\nonumber {} & \hspace{.2in} + O\left(\frac{1}{k^{2/p}}\cdot
\left(\frac{1}{T^{2t-1}} - 1\right)\cdot \frac{1}{\frac 1t - 2}\right)
+ O(2^{-\Omega(k)}\cdot (T - 1))\\
\nonumber {} & = \E[f(R)] + O(\eps)
\end{align}

We show $\E[f^2(R)] = O(|Q|^{2/t})$ similarly.  Namely, we argue
through the chain of inequalities
$$ \E[|Q|^{2/t}] \approx_{\eps} \E[f^2(Q)] \approx_\eps
\E[f^2(R)] ,$$
which proves our claim since $\E[|Q|^{2/t}] = \Omega(1)$ by
\Theorem{zolotarev}.

\vspace{.1in}

\noindent $\mathbf{\E[|Q|^{1/t}] \approx_{\eps}
\E[f^2(Q)]}$: We have
\begin{align*}
 |\E[|Q|^{2/t}] - \E[f^2(Q)]| & = 2\int_{T^t}^\infty (x^{2/t} - T^2)\cdot
\varphi_p(x)dx \\
{} & = \int_{T^t}^\infty (x^{2/t} - T^2)\cdot O(1/x^{p+1})dx\\
{} & = O\left(T^{2 - tp}\cdot \left(\frac{t}{pt-2} -
    \frac{1}{p}\right)\right)\\
{} & = O(1/(Tp))\\
{} & = O(\eps)
\end{align*}

\vspace{.1in}

\noindent $\mathbf{\E[f^2(Q)] \approx_{\eps}
\E[f^2(R)]}$: This is argued nearly identically as in our proof that
$\E[f(Q)] \approx_\eps \E[f(R)]$ above. The difference is that our
error term now corresponding to \Equation{geometric-error} is
\begin{align*}
 \int_0^1x^{2/t - 1} & \cdot O(k^{-1/p} +
k^{-2/p} + 2^{-\Omega(k)}))dx +
\int_1^{T^t}x^{2/t - 1}\cdot O(k^{-1/p}/x^{p+1} + k^{-2/p}/x^2
+ 2^{-\Omega(k)}))dx\\
{} & = O(\eps) + O\left(\frac{1}{k^{1/p}}\cdot
\left(\frac{1}{T^{tp + t-2}} - 1\right)\cdot \frac{1}{\frac 2t - p
  -1}\right)\\
{} & \hspace{.2in} + O\left(\frac{1}{k^{2/p}}\cdot
\left(\frac{1}{T^{2t-2}} - 1\right)\cdot \frac{1}{\frac 2t - 2}\right)
+ O(2^{-\Omega(k)}\cdot (T^2 - 1))\\
{} & = O(\eps)
\end{align*}
\end{proof}

\end{document}